\documentclass[aps,prd,groupedaddress,showpacs,nofootinbib,amssymb,balancelastpage,preprintnumbers]{revtex4}

\usepackage{tabularx}

\usepackage[dvipdfmx]{graphicx}
\usepackage{graphicx,bm,color}
\usepackage{amsmath}
\usepackage{amssymb}
\usepackage{amsfonts}
\usepackage{cases}
\usepackage{cancel}
\usepackage{hyperref}
\usepackage{ulem}

\usepackage{here}

\def\ra{\rightarrow}
\def\L{\left(}
\def\R{\right)}
\def\wt{\widetilde}

\def\ld{\lambda}
\def\f{\frac}
\newcommand{\be}{\begin{equation}}
\newcommand{\ee}{\end{equation}}
\newcommand{\bea}{\begin{eqnarray}}
\newcommand{\eea}{\end{eqnarray}}
\newcommand{\ba}{\begin{array}}
\newcommand{\ea}{\end{array}}

\long\def\symbolfootnote[#1]#2{\begingroup%
\def\thefootnote{\fnsymbol{footnote}}\footnote[#1]{#2}\endgroup}

\newcommand{\beq}{\begin{equation}}
\newcommand{\eeq}{\end{equation}}

\newcommand{\gev}{\, {\rm GeV}}

\begin{document}

\title{Dark Confinement-Deconfinement Phase Transition: \\
A Roadmap from Polyakov Loop Models to Gravitational Waves}

\author{Zhaofeng Kang}
\email[E-mail: ]{zhaofengkang@gmail.com}
\affiliation{School of physics, Huazhong University of Science and Technology, Wuhan 430074, China}

\author{Shinya Matsuzaki}
\email[E-mail: ]{synya@jlu.edu.cn}
\affiliation{Center for Theoretical Physics and College of Physics, Jilin University, Changchun, 130012,
China.}

\author{Jiang Zhu}
\email[E-mail: ]{jackpotzhujiang@gmail.com}
\affiliation{School of physics, Huazhong University of Science and Technology, Wuhan 430074, China}

\date{\today}

\begin{abstract}

We explore the confinement-deconfinement phase transition (PT) of the first order (FO) arising in $SU(N)$ pure Yang-Mills theory, 
based on Polyakov loop models (PLMs), in light of the induced gravitational wave (GW) spectra.  We demonstrate that the PLMs with the Haar measure term, involving  models successful in QCD with $N=3$,  are potentially incompatible with the large $N$ scaling for the thermodynamic quantities and the latent heat at around the criticality of the FOPT  reported from the lattice simulations.   We then propose a couple of models of polynomial form, which we call the 4-6 PLM (with four- and six-point interactions among the basic PL fields which have center charge 1) and 4-8 PLM (with four- and eight-point interactions),  and discuss how such models can naturally arise in the presence of a heavy PL with charge 2. We show that  those models give the consistent thermodynamic and large $N$ properties at around the criticality. The predicted GW spectra are shown to have high enough  sensitivity to be probed in the future prospected interferometers such as BBO and DECIGO.

\end{abstract}


\maketitle
\section{Introduction}

Pure Yang-Mills (PYM) sector surviving at low energy is of particular interest in the new physics domain, in the context of dark sector of the universe, including 
dark matter (see, e.g., a review~\cite{Kribs:2016cew} and also~\cite{Yamanaka:2019aeq,Yamanaka:2019yek,Yamanaka:2019gak})   
and also in the string theory~\cite{string}. 
However, if such a PYM sector is sufficiently secluded from the visible sector, then what is its detectable signal?

Due to the asymptotic freedom of non-Abelian gauge theory, the PYM theory in the ultraviolet scale lies in the deconfinement phase described by ``free" gluons, whereas flowing down to the infrared scale, the interactions between gluons steadily get strong and eventually confine themselves to the colourless glueballs~\footnote{These objects may be stable or sufficiently long-lived, providing a good candidate for the feebly interacting massive particle as a dark matter~\cite{Kang:2019izi}. In that case, the glueball decays and may leave detectable signals in the cosmic rays.}. Therefore, there is a confinement-deconfinement transition at the critical temperature $T_c$. 
It is well-known that for the $SU(N)$ PYM theory with $N>2$, this phase transition (PT) is first order (FO). Then, one expects the corresponding FO cosmic/thermal PT in the early universe, during which gravitational waves (GWs) are produced. They immediately decouple from the matters and are wandering in the universe today, which may meet the GW detectors in the sky such as 
the LISA~\cite{Audley:2017drz,Baker:2019nia}, 
Big Bang Observer (BBO)~\cite{Crowder:2005nr,Corbin:2005ny,Harry:2006fi,Thrane:2013oya,Yagi:2011wg}, DECIGO~\cite{Yagi:2011wg,Seto:2001qf,Isoyama:2018rjb} and TianQin~\cite{Lu:2019sti},  where the sensitivities of frequencies cover a wide region from $10^{-3}$mHz to kHz; and the sensitivities can be improved in orders of magnitude  over the  next three decades.  


To quantify the GW prediction, working on phenomenological models of Ginzburg-Landau type modeling the confinement-deconfinement PT is helpful. 
Since the PT is subject to the nonperturbative aspect of the PYM theory, 
the output from lattice simulations is often utilized to extract important clues in constructing the phenomenological models. 
Lattice simulations present the equilibrium thermodynamic properties of the $SU(N)$ plasma at finite $T$, in particular for $N>3$~\cite{Borsanyi:2012ve,Lucini:2003zr,Lucini:2005vg,Panero:2009tv,Datta:2010sq,Lucini:2012wq}.  The pressure $p(T)$ is directly related to the potential of the model in terms of some fundamental degrees of freedom such as the Polyakov loop (PL)~\cite{PL}, thus it is of special importance: 

\begin{itemize}

\item   The pressure is found to scale as the ideal Stefan-Boltzmann (SB) limit when $T>4T_c$, namely $p(T)\sim (N^2-1)T^4$. On the contrary, the pressure tends to vanish below $T_c$.

\item A feature is observed by Meisinger and Pisarski etc~\cite{RMM,EFT:1}~\footnote{It is first based on the precise data for $N=3$~\cite{feature1,Lucini:2005vg,Borsanyi:2012ve}, but a similar feature is also shown to be present for different $N$ as well~\cite{Lucini:2003zr,Lucini:2005vg,Panero:2009tv,Datta:2010sq,Lucini:2012wq} }: The conformal anomaly $\Delta(T)$, the energy density $\epsilon(T)$ minus $3p(T)$, is nearly a constant relative to the gluon gas limit within the semi-quark gluon plasma (sQGP) region $1.2T_c<T<4T_c$. This suggests that the nonperturbative correction may scale as $T^2$, and consequently in the sQGP region the pressure takes the form 
\begin{equation}\label{SQGP}
\begin{aligned}
p(T)=c_1(T^4-c_2T_c^2T^2),
\end{aligned}
\end{equation}
with $c_2\simeq 1$. The first term is supposed to be obtained after integrating out the hard modes at ${\cal O}(T)$ in the weak-coupling expansion, recovering the ideal gluon gas limit indicated above. In terms of the PL, $c_1$ is determined by the perturbative Weiss potential at leading order~\cite{WPT}. 
To properly match with lattice data from $T_c$ to $1.2T_c$, however, 
it is found that other terms such as the linear term $T_c^3T$ become important. 
Without a convincing  theory to justify Eq.~(\ref{SQGP}) in the whole region of $T>T_c$, one should keep one's mind open. We will come back to this point 
later. 

\end{itemize}

Besides, the latent heat released during the FOPT shows a simple scaling behavior, i.e.,  $L_N\approx 0.388(N^2-1)T_c^4$~\cite{Borsanyi:2012ve,Lucini:2003zr,Lucini:2005vg,Panero:2009tv,Datta:2010sq,Lucini:2012wq}. 
Then, following the Polyakov loop model (PLM), one should find an effective potential of the PL encoding these features.

To this end, first of all, the potential of the PLM at hand should realize a FOPT, which severely limits the profile of the potential. To our knowledge, there are several types of PLM available which give rise to the FOPT supported from the lattice simulations. 
One type makes the FOPT triggered by the Haar measure term (the Vandermonde determinant interaction term) associated with the $SU(N)$ group integral over the PL variable~\cite{Haar0}. 
Including a quadratic term of the PL variable (corresponding 
to the nearest-neighbor kinetic interaction term~\cite{Haar0}), 
this Haar measure approach is shown to be consistent with the picture of ghost dominance in the infrared regime~\footnote{The ghost dominance is associated with the Kugo-Ojima scenario of confinement, realized by some nonperturbative propagators of the ghost and the gluons~\cite{Braun:2007bx,Pawlowski}, leading to the inverted Weiss potential.}, and is indeed successful in the case of QCD with 
$N=3$ even with quarks. 
We shall label this approach as ``Haar-type". 
Based on this Haar measure prescription, Ref.~\cite{Kubo:2018vdw} investigated a dark $SU(N)$ gauge theory with colored scalars (in a scale-invariant setup) and the FOPT for the confinement-deconfinement, and discussed the related GW signals. However, the Haar measure term will be almost impossible to handle as $N$ becomes large. Moreover, as will be clarified in the present paper, this Haar-type model suffers from incompatibility with the lattice data for $N=4,5,6$ in addressing the large $N$ scaling for the thermodynamic quantities at around $T=T_c$.

As another approach, there is what is called the matrix model, motivated by the form of the Weiss potential~\cite{RMM}.  This model treats the eigenvalues of the PL as variables~\cite{RMM,Dumitru:2012fw}, where the FOPT is understood by the mutual repulsion of these eigenvalues 
~\cite{PLM1,Dumitru:2001xa}.  
Very recently, Ref.~\cite{Halverson:2020xpg} appeared which studies GWs from the PYM sector following the matrix-model approach~\footnote{We will not cover this approach in the present paper, but just give some comments on comparison in the last section, and will 
leave deeper relationship with what we will propose in another publication.   
}.

Though there are several models in the ballpark, it is still unclear how to realize the  FOPT of the confinement-deconfinement for a generic $N$, consistently with the large $N$-thermodynamic feature relevant to the GW spectral predictions. In this paper, we explore $SU(N)$ PLMs which can fully be consistent with the large $N$ scaling for the thermodynamic quantities at around the criticality of the confinement-deconfinement PT, including the aforementioned thermodynamic properties above, hence can provide the proper GW spectral signals today.  We demonstrate that the Haar-type model is potentially incompatible with those critical-large $N$ scaling -properties reported from the lattice simulations.

 We then propose a couple of simple-minded polynomial models, which we call the 4-6 PLM (having four- and -six point interaction terms) and  4-8 PLM (with four- and -eight point interaction terms),  based on the spirit outlined in the literature~\cite{Pisarski:2001pe} long ago.  We show that  the 4-6 and 4-8 PLMs can fully be consistent with the required thermodynamic and large $N$ properties.

Actually, those PLMs are  not completely novel, but have not received sufficient attention yet. 
The present work provides a possible derivation of the 4-6 PLM as well as the 4-8 PLM, and for the first time thoroughly analyzes the models, in light of GW spectra taking into account consistency with the large $N$ scaling coupled to the criticality of the confinement-deconfinement FOPT. We show that the predicted GW spectra can have high enough sensitivity to be probed in the future prospected interferometers such as BBO and DECIGO.

The paper is organized as follows: In Section II we start with a review of the generic ingredients to 
construct the PLMs including the concept of $Z_N$ center symmetry, the order parameter of the confinement-deconfinement PT, 
and introduction of a nonminimally-$Z_N $charged PL variable (part A). 
With these preliminaries at hand, 
we employ a widely used conventional PLM, called the Haar-type PLM, and show that the model turns out to have incompatibility in a sense of the large $N$ scaling for the latent heat and the other thermodynamic quantities at around the criticality for the confinement-deconfinement PT (part B). 
Then, we propose a set of PLMs which properly realize the thermodynamic and the large $N$ properties at around the criticality, that is what we call the 4-6 and 4-8 PLMs, with explanation on the derivation for those highly nonminimal coupling terms. 
The models are shown to yield the good-fitness with the $SU(N)$ lattice data on the thermodynamic quantities, as well as the latent heat, regarding the large $N$ scaling around the criticality (part C).

In Sec.III, we move on to discussion on the generation mechanism of the GW spectra by the FOPT based on the 4-6 and 4-8 PLMs. The dark $SU(N)$ PYM is assumed to be secluded-dark gluonic plasma (part A). Passing discussion on a couple of the well-known GW sources (part B), we numerically evaluate the GW spectra, in comparison with the prospected GW detector sensitivities (part C). Our conclusion and several issues still left with us are presented in Sec.IV.

\section{Road to a Proper PLM}

\subsection{PLs and $Z_N$ center symmetry}

The PL plays a center role in understanding the  confinement-deconfinement PT in the PYM theory~\cite{PL}. 
When a immovable test quark is placed at the position $\vec{x}$, its color-averaged free energy $F_q(\vec{x})$ can be expressed as $e^{-F_q(\vec{x})/T}=\langle{\rm tr}_c L(\vec{x})\rangle$ with $L(\vec{x})$ the thermal  Wilson line at a fixed spatial position $\vec{x}$ extending in the temporal direction: 
\begin{equation}\label{PL:def}
\begin{aligned}
L(\vec{x})={\cal P}\exp{\left[ig\int_0^{1/T}dx_4 A_4(\vec{x},x_4)\right]},
\end{aligned}
\end{equation}
where $g$ denotes the PYM gauge coupling, and the path integration interval is closed due to periodicity of the gauge field $A_\mu$: 
$A_4(\vec{x},x_4)=A_4(\vec{x},x_4+1/T)$. Here $A_4\equiv A_4^a t^a$ with $t^a$ $(a=1,\cdots , N^2-1)$ being the generators of $SU(N)$ in the fundamental representation. Under the local $SU(N)$ transformation, $L$ transforms as the adjoint representation. Therefore, taking trace in the color space, one obtains a gauge invariant operator, that is the PL: 
\begin{equation}\label{}
\begin{aligned}
{l}(\vec{x})=\f{1}{N}{\rm tr}_c L(\vec{x}) . 
\end{aligned}
\end{equation}
Quantum mechanically, ${l}(\vec{x})$ is a local operator, while classically it is a complex scalar field with the field value bounded by one: 
$|l(\vec{x})|\le 1$. 
Note that ${l}(\vec{x})$ is dimensionless by construction, and one should make up it by some dimensionful quantity in addressing the cosmic/thermal evolution of the GWs; we will come back to this point in Sec.~\ref{GW} (around Eq.(\ref{EOM})).


Due to the compactified temporal direction,  gauge fields should respect the periodicity shown before, however, gauge transformations $V(\vec{x},x_4)$ can be non-periodic. Instead, to keep the PYM action gauge invariant, they are just required to satisfy the twisted boundary condition (the center twisted gauge transformations): $V(\vec{x},\beta)=\wt z_kV(\vec{x},0)$, where $\wt  z_k=e^{i2k\pi /N }$ with $k=1, \cdots ,N$ belonging to the discrete group $\wt Z_N$; the transformation satisfying the periodicity corresponds to  $k=N$. In particular, the gauge transformation with respect only to $x_4$ 
\begin{equation}\label{}
\begin{aligned}
V_k(x_4)=\L z_k\R^{x_4/\beta}=e^{i2k\pi x_4 /(N\beta)} \cdot 1_{N\times N}
\end{aligned}
\end{equation}
obviously obeys  the twisted boundary condition; it is dubbed the center symmetry~\cite{center} since it is simply the power of the elements of $z_k=\wt z_k 1_{N\times N}\in Z_N$, the center of $SU(N)$. Under the action of  $V_k(x_4)$, 
${l}$  is not invariant  and changes as ${l}\ra z_k {l} $ 
although the Euclidean action is invariant. 
In this sense, ${l}$ is charged under the global center $Z_N$. Now,  it is seen that the gauge invariant operator ${l}$ can serve as a good order parameter of the confinement-deconfinement PT:
\begin{itemize}
\item  If  $\langle{\rm tr}_c L(\vec{x})\rangle=0$ and $Z_N$ is not broken, then $F_q\ra \infty$, thus quarks are confined.
\item   If  $\langle{\rm tr}_c L(\vec{x})\rangle\neq 0$ and $Z_N$ is broken, then $F_q=$ finite, thus quarks are deconfined. 
 \end{itemize}

The exact $Z_N$ charge of ${l}$ is not determined from the first principle, and one may specify its charge as 1 and dub it as the fundamental PL, $l_1 \equiv 
\frac{1}{N}{\rm tr}(L)$. 
Other traced PLs with different $Z_N$ charges can be introduced~\cite{Pisarski:2000eq,Pisarski:2001pe}. Or, we introduce traced PLs in a different representation $R$ under $SU(N)$, normalized by the dimension $d_R$. But different representations may have the same $Z_N$ charge. For instance, in the large $N$-limit, both PLs in the representations of $d_{R,\pm}=(N^2\pm N)/2$ have charge 2. In the large $N$ limit, according to group theory~\cite{decompose}, they can be decomposed as
\begin{equation}\label{ZN:other}
\begin{aligned}
{l}_{2,\pm}={l}_1^2\pm \f{1}{N^2}{\rm tr}L^2\,, 
\end{aligned}
\end{equation}
where tr$L^2$ means the Wilson line wrapped around the imaginary time forward twice, and it is negligible in the large $N$ limit. In general, PLs with $Z_N$ charge $p_+-p_-$ (module $N$) can be expanded as: 
\begin{equation}\label{ZN:other1}
\begin{aligned}
\sum c_{n,m}(N)l_1^{p_+-m}{\rm tr}L^{m}\bar l_1^{p_--n}{\rm tr}(L^{\dagger})^{n}, 
\end{aligned}
\end{equation}
with $\bar l_1\equiv {\rm tr}(L^\dagger)/N$, the PL in the anti-fundamental representation. To determine $c_{n,m}(N)$ is nontrivial~\cite{decompose,Dumitru:2003hp}, but irrelevant to our discussions. In constructing PLM incorporating additional PLs, we just treat them as independent degrees of freedom, and require that the couplings among them are constrained by $Z_N$. The higher-$Z_N$ charged PLs are assumed to be heavy and can be integrated out, which leads to the  polynomial potential that we will use 
in Sec.II (See Eqs.(\ref{U-l1-l2}) and (\ref{Ul-eff}), and discussions around there.).  


\subsection{The Haar-type PLMs for general $SU(N)$ theory}

\subsubsection{The models}

 The Haar-type PLM is constructed based on the two-parameter model proposed by Fukushima~\cite{Haar0,Haar1} for QCD with $N=3$. It contains the nearest-neighbor coupling among PLs, so-called the kinetic term for the traced PL $l$, and as well as the Haar measure term $H_N[L]$ which is crucial to trigger the FOPT. The Haar measure incorporates $(N-2)$ degrees of freedom other than $l$, which causes the complexity for a larger $N$. In the Polyakov gauge where $A_4$ is diagonal,
\begin{equation}\label{}
\begin{aligned}
A_4(\vec{x})=\f{2\pi}{g\beta}{\rm diag}(q_1(\vec{x}),q_2(\vec{x}),....,q_N(\vec{x}))
\end{aligned}
\end{equation}
with $\sum_iq_i=0$, then $H_N[L]$ is given by
\begin{equation}\label{}
\begin{aligned}
H_N[L(\vec{x})]=\Pi_{i<j}|e^{i2\pi q_i(\vec{x})}-e^{i2\pi q_j(\vec{x})}|^2.
\end{aligned}
\end{equation}
The Haar-type PLM, generalized from the original model for $N=3$~\cite{Haar0,Haar1} to any color~\cite{Kubo:2018vdw}, thus goes like 
\begin{equation}\label{}
\begin{aligned}
\frac{V(L,T)_{\rm Haar,0}}{T^4}=&\mathcal{V}_{\rm Haar,0}(L,T)=C_2(N)\left[6 \, {\rm exp}(-C_1(N)/T)N^2l l^\dagger+\log{H_N[L]}\right],
\end{aligned}
\end{equation}
where the form of the kinetic term has been fixed inspired by the strong coupling expansion; $C_1(N)$ and $C_2(N)$ are two parameters needed to be determined by lattice data. Although it successfully explains the confinement-deconfinment PT of the first order, 
it can not reproduce proper thermodynamics observed in lattice simulations, as 
will be clarified later. 

Soon later we will show that it is true even in another Haar-type PLM (for $N=3$) studied in Ref.~\cite{para2}, which has a kinetic term including more parameters, generalized to any color number:
\begin{equation}\label{Potential}
\begin{aligned}
\mathcal{V}_{\rm Haar,1}(L,T)=&-\frac{a(T)}{2}l l^\dagger+b(T)\log{H_N[L]}
, \\
a(T)=&a_0+a_1 \left(\frac{T_c}{T}\right)+a_2\left(\frac{T_c}{T}\right)^2\ \ \ \ b(T)=b_3\left(\frac{T_c}{T}\right)^3.
\end{aligned}
\end{equation} 
The potential parameters $a_0, a_1, a_2$ and $b_3$ are constrained and fixed by imposing some physical conditions and fitting this potential to the lattice data on the thermodynamic quantities. 

\subsubsection{Constraining the potential parameters from first-order deconfinement PT}

As to the required thermodynamic constraints, there are two. 
One is set by requiring that as noted in the Introduction, 
at high temperature the PL potential should realize the SB limit in the deconfinement vacuum $l=l_c$, which must go to 1 as $T\rightarrow \infty$ (because in this limit $L\rightarrow 1$ by definition, see Eq.~(\ref{PL:def})) and therefore $T^4 \mathcal{V}(T)=-\frac{(N^2-1)\pi^2}{45}T^4$. It fixes $a_0$ in Eq.(\ref{Potential}):
\begin{align} 
a_0=\frac{2(N^2-1)\pi^2}{45}. 
\label{a0}
\end{align}The other condition is given by assuming that the FOPT takes place at the critical temperature $T_c$, namely that two minima should be degenerate at $T=T_c$. This condition gives a set of equations 
\begin{equation}\label{GEQ}
\begin{aligned}
&\mathcal{V}(L_c,T_c)=\mathcal{V}(L_0,T_c)
, \\
&\frac{\partial\mathcal{V}(L,T)}{\partial L} \Bigg|_{L=L_c,T=T_c}=0
, \\
&\frac{\partial\mathcal{V}(L,T)}{\partial L} \Bigg|_{L=L_0,T=T_c}=0
, 
\end{aligned}
\end{equation}
where the labels $c$ and $0$ attached on the $L$s respectively stand for  the values measured in the deconfinement phase ($T>T_c$) and the confinement phase ($T<T_c$). The last condition in Eq.(\ref{GEQ}) gives the stationary point in the confinement phase, which has to be realized at $L_0 = l_0 = 0$ so as to reflect the $Z_N$ center symmetry in the confinement phase. Note from the potential form in Eq.(\ref{Potential})  that this condition is controlled solely by the Haar measure term, no matter what values independent of the potential parameters $a_0,a_1,a_2,b_3$ take.  The remaining first two conditions in Eq. (\ref{GEQ}) read 
\begin{equation}
\begin{aligned}
&-\frac{a_0+a_1+a_2}{2}l_c l_c^\dagger+b_3\log{H_N(L_c)}=b_3\log{H_N(L_0)}
, \\
&-(a_0+a_1+a_2)l_c+\frac{b_3}{H_N(L_c)}\frac{\partial H_N(L_c)}{\partial l_c}=0, 
\end{aligned}
\end{equation}
i.e.,  
\begin{equation}\label{EQV}
\begin{aligned}
&b_3= l_c \frac{H_N(L_c)}{\frac{\partial H_N(L_c)}{\partial l_c}} (a_0+a_1+a_2)
, \\
&-\frac{l_c^\dagger}{H_N(L_c)}\frac{\partial H_N(L_c)}{\partial l_c} +2\log{H_N(L_c)}=2\log{H_N(L_0)}
. 
\end{aligned}
\end{equation}
Thus, given the Haar measure term $H_N$, the field value $l_c$ can be numerically determined by solving the second relation in Eq.(\ref{EQV}), as is independent of the potential parameters.  With $l_c$, we combine the first and second relations in Eq.(\ref{EQV}), and obtain the relation between the potential parameters
\begin{equation}
b_3= - \frac{l_c^\dagger l_c}{2}\frac{1}{\log[H_N(L_0)/H_N(L_c)]}(a_0+a_1+a_2).
\label{b3-condition}
\end{equation} 

\subsubsection{Fitting the potential parameters from thermodynamics }

The potential parameters should also be fitted to the thermodynamic quantities measured in lattice simulations, such as pressure $P$, energy density $\epsilon$ and the entropy density $s$. From the potential in Eq.(\ref{Potential}), those thermodynamic quantities  can be evaluated as 
\begin{equation}\label{TF}
\begin{aligned}
P(T)&=-V(T)
, \\
\epsilon(T)=\frac{dU(T)}{dV}&=T\frac{dS(T)}{dV}-P(T)=T\frac{dP(T)}{dT}-P(T)
, \\
s(T)=\frac{dS(T)}{dV}&=\frac{P(T)+\epsilon(T)}{T}
. 
\end{aligned}
\end{equation}
In light of GW signals, of particular importance among the thermodynamic properties is the latent heat for the confinement-deconfinement PT of the first order.  Given the PL potential as in Eq.(\ref{Potential}),  the latent heat $L_N$ for $SU(N)$ PYM theory is defined as  
\begin{equation} 
L_N(T_c)=-\Delta V(T_c)+T\frac{\partial \Delta V(T)}{\partial T} \Bigg|_{T=T_c},
\label{LN-def}
\end{equation}
where $\Delta V(T)=V(L_0,T)-V(L_c,T)$. Taking into account the first relation in Eq.(\ref{GEQ}), we have
\begin{equation}
\begin{aligned}
L_N&=T_c\frac{d V(L_0,T)}{d T} \Bigg|_{T=T_c}-T_c\frac{d V(L_c,T)}{d T} \Bigg|_{T=T_c}
\\
&=T_c^4\left[(4 a_0+3 a_1+2 a_2)\frac{l_c l_c^\dagger}{2}+b_3\log \frac{H_N(L_0)}{H_N(L_c)}\right] 
. 
\end{aligned}
\end{equation}
To this $L_N$, lattice simulations tell us~\cite{Borsanyi:2012ve,Lucini:2003zr,Lucini:2005vg,Panero:2009tv,Datta:2010sq,Lucini:2012wq} 
\begin{equation}\label{Lat}
\frac{ L_N}{(N^2-1)T_c^4} \simeq 0.388-\frac{1.61}{N^2}
, 
\end{equation}
which implies 
\begin{equation}\label{LHC}
\begin{aligned}
\log\left[ \frac{H_N(L_0)}{H_N(L_c)} \right]\cdot b_3 
\simeq (N^2-1)\left(0.388-\frac{1.61}{N^2}\right)-\frac{l_c l_c^\dagger}{2}(4 a_0+3 a_1+2 a_2). 
\end{aligned}
\end{equation}
This is the potential parameter relation for the Haar-type PLM,  which must be satisfied to be consistent with the vicinity of the FOPT in the $SU(N)$ PYM theory. Actually, using $b_3$ given in Eq.~(\ref{b3-condition}), the above equation turns out to be a relation among $a_i$:
\begin{equation}
\frac{l_c l_c^\dagger}{2}(3a_0+2a_1+a_2) \simeq 
(N^2-1)\left(0.388-\frac{1.61}{N^2}\right),
\end{equation}
where $a_0$ has been fixed as in Eq.(\ref{a0}).

It turns out, however, that the potential parameters in the Haar-type PLMs are inconsistent with the lattice data on thermodynamic quantities in $SU(N)$ gauge theory, in terms of the large $N$ scaling. We come to this conclusion with great help from the work of Kubo~\cite{Kubo:2018vdw}, and let us explain the details in the following.

\subsubsection{No Go in a View of Large $N$ Scaling and the SU(N) PYM Criticality}

From the numerical result~\cite{Kubo:2018vdw} that no matter how large $N$ is taken, $l_c$ is found to be around 0.5.  Taking into account this result and $a_0=\frac{2(N^2-1)\pi^2}{45}$ in Eq.(\ref{a0}), we have 
\begin{equation}
\frac{1}{8}(2a_1+a_2) \simeq (N^2-1)\left(0.224-\frac{1.61}{N^2} \right)\,. 
\end{equation}
Thus the parameters $a_0, a_1$ and $a_2$ can scale with $(N^2-1)$ for a sufficiently large $N$. On the contrary, actually, the parameter $b_3$ cannot do because of 
nontrivial $N$ dependence of the Haar measure part present as a prefactor 
in Eq.(\ref{b3-condition}). 
As references,  
we know~\cite{Kubo:2018vdw} that $H_3(L_0)/H_3(L_c)\sim0.9,\ H_4(L_0)/H_4(L_c)\sim2.6,\ H_5(L_0)/H_5(L_c)\sim4.8$, and $H_6(L_0)/H_6(L_c)\sim12.7$, which imply 
\begin{equation}
\frac{\log [H_{N_1}(L_0)/H_{N_1}(L_c)]}{\log [H_{N_2}(L_0)/H_{N_2}(L_c)] }>\frac{N_1^2-1}{N_2^2-1}
, 
\end{equation}
for large enough $N_1$ and $N_2$. 
Therefore, the parameter $b_3$ cannot simply scale with $(N^2-1)$. 
This is in contradiction with  
the lattice result~\cite{Datta:2010sq} showing the universal large $N$ scaling for 
all thermal quantities $\propto (N^2 -1)$, 
which indicates that all Polyakov loop potential parameters should be proportional to $N^2-1$ as well. 
This incompatibility arises from the Haar measure part included in 
the potential, driving the nontrivial $N$ dependence into the thermodynamics 
in $SU(N)$ gauge theory.

As a concrete example, let us take a look at the case with $N=4$. 
Then we have
\begin{equation}\label{SU4P}
\log\frac{H_4(L_0)}{H_4(L_c)} \simeq 2.5562
. 
\end{equation} 
Hence Eq.~(\ref{LHC}) reads  
\begin{equation}
2.5562\times b_3 \simeq 4.311-0.1247\times(4a_0+3a_1+2a_2)
. 
\end{equation}
On the other hand, from Eq.(\ref{b3-condition}), we have 
\begin{align}
b_3 \simeq -0.0488(a_0+a_1+a_2) 
. 
\end{align}
Combining them gives 
\begin{equation}
a_2 \simeq 34.5562-(3a_0+2a_1). 
\end{equation} 
Substituting these conditions into the fitting program for the thermodynamic quantities in $SU(4)$ gauge theory, with data available from~\cite{Datta:2010sq}, we have found the fit does not match the lattice result at all~\footnote{
We have checked that this failure will not be cured even removing the SB limit condition in Eq.(\ref{a0}) which 
has not well been reproduced in lattice simulations somehow. 
}.

Thus, it is concluded that the Haar type PL potential in Eq.(\ref{Potential}) for generic $SU(N)$ gauge theory possesses incompatibility between the latent heat, i.e., confinement-deconfinement phase transition feature, and the overall large $N$ scaling for the thermodynamic quantities.

\subsection{A Proper Model: Polynomial Potential}

\subsubsection{The 4-6 PLM versus 4-8 model}

Now that we have demonstrated the Haar type PLM in Eq.(\ref{Potential}) to have incompatibility between the latent heat nature of the FOPT and the large $N$ scaling, we need to consider other PLMs. For $N=3$, there is a very successful potential structure of polynomial form with powers of ${l}$ not exceeding 4 (renormalizablility)~\cite{Ratti:2005jh}, where the FOPT is triggered by the well-known cubic term $\L{l}^3+{l}^{*3}\R$ by virtue of the underlying $Z_3$~\cite{center}. Nevertheless, this minimal polynomial form obviously fails for a generic $Z_N$. On the other hand, the large $N$ scaling property revealed by lattice data strongly indicates that there must be a simple way to construct the PLM, irrespective of $N$ not less than 3.

Thus we propose a simple PLM containing three terms, 
\begin{equation}\label{PP}
\mathcal{U}(l,T)=\frac{V(l,T)}{T^4}=-\frac{a(T)}{2}|l|^2+b|l|^4+c|l|^6
, 
\end{equation}
All the terms in Eq.~(\ref{PP}) are allowed by $Z_N$ symmetry, and even 
keep a global $U(1)$ symmetry. 
Actually, it turns out that this enhanced $U(1)$ symmetry allows us to derive a unified large $N$ scaling property of the PLM. 
Hence we will ignore other terms that may be allowed by the $Z_N$ symmetry, for instance $l^4+l^{*4}$ for $N=4$, and so on.

Comments on the potential parameters in Eq.(\ref{PP}) are in orders. It is assumed that $a(T)=a_0+a_1(\frac{T_c}{T}) +a_2(\frac{T_c}{T})^2+a_3(\frac{T_c}{T})^3$ as usual, but $b$ and $c$ are $T$-independent. In order to fit the sQGP region $1.2T_c<T<4T_c$, the pressure may take the simple form shown in Eq.~(\ref{SQGP}), and consequently the $a_1$-term namely the  $T^3$ term (in terms of the potential $V(l,T)$) 
is not important while the $a_2$-term is important. 
However, the $a_3$-term namely the $T$ term is crucial to fit the region very close to $T_c$, which will be confirmed by fitting to the lattice data later (See Figs.~\ref{fitting-plots}, 
\ref{4-8:PLM} and Table~\ref{table-1}). 
A different viewpoint has been proposed by Fukushima in Ref.~\cite{Haar1}: 
The PL potential is merely dominant to the contributions to the pressure of the system below about $2T_c$, and it should give way to the transverse gluons at  higher temperatures; therefore, the PLM giving a pressure below the lattice data  for $T \gtrsim 2T_c$ is acceptable. We will investigate both scenarios (with or without $a_2$) 
in the lattice fitting procedure, and find that, as expected, they are almost ``degenerate'' PLMs in a view of the  GW signals (See Fig.~\ref{GW:6-8}).


Our PLM is characterized by two features: 1) The quartic term with a negative sign $b<0$; 2) a nonrenormalizable term with a positive sign $c>0$ to stabilize the potential. These two terms create the deconfined vacuum at ${l}\neq 0$. In the mean field theory, the  nonrenormalizable terms usually are not taken into account, but its legitimacy, along with the sign of the quartic term, may be justified in the presence of additional and non-minimally charged PLs defined in Eq.~(\ref{ZN:other}). This potential form frees from extra large $N$ scaling, other than the overall factor of $(N^2-1)$ for all the potential parameters at around the $SU(N)$ gauge criticality, as will be clearly seen below.   


Let us briefly explain  how the 4-6 structure of the PLM in Eq.(\ref{PP}) can arise following the spirit of Pisarski outlined in Ref.~\cite{Pisarski:2001pe}. The key is the presence of charge-two PL,  ${l}_2$, which always allows a cubic coupling with the fundamental PL, $l$, like  ${l}^2{l}_2^*$. The most general two-field model is given by
\begin{equation}\label{U-l1-l2}
\mathcal{U}({l},{l}_2)\supset {m_2^2}|{l}_2|^2+m^2|{l}|^2+ \left[\ld_1(ll^*)^2+\ld_2({l}_2{l}_2^*)^2+\ld_{12}|l|^2|l_2|^2\right]+ a_{12}\L{l}^2{l}_2^*+c.c.\R,
\end{equation}
with taking $a_{12}$ to be real. Assuming $m_2$ to be heavy while taking $m$ to be light around $T_c$, then ${l}_2$ can be integrated out. Among others, this procedure produces four-point interactions $|{l}_1|^4$ with a negative coupling $-a_{12}^2/m_2^2\equiv -\wt \ld_2<0$. 
The negativeness of the induced quartic coupling is robust, because, in terms of the $l l \to ll$ scattering process, 
it arises from the $l_2$-scalar exchange   
involving the trilinear $a_{12}$ interaction, which 
gives rise to the repulsive potential among the two $l$ probes (in the nonrelativistic limit). 
The total effective PLM below the $m_2$ scale reads 
\begin{equation}\label{Ul-eff}
\mathcal{U}({l})_{eff}\supset m^2|{l}|^2+ \L\ld_1-\wt \ld_2\R |{l}|^4 +\ld_{12}\wt \ld_2^2|{l}|^6+\ld_2 \wt \ld_2^2|{l}|^8.
\end{equation}
Hence, if the bare coupling $\ld_1$ is properly chosen to make $\ld_1<\wt\ld_2$, the total quartic coupling is naturally negative, triggering the FOPT. Moreover, when we turn off either $\ld_{2}$ or $\ld_{12}$ for simplicity or minimality,  then the above model respectively gives rise to the 4-6 (as in Eq.(\ref{PP})) or 4-8 structure as desired -- which we shall call the 4-6 and 4-8 PLMs, respectively.

The case $N=2$ is an exception since the confinement-deconfinement PT becomes second order, which may be accidentally due to the smallness of $\wt \ld_2$. Actually, in this case ${l}_2$ is the ``quarkless-baryon" loop~\cite{Pisarski:2001pe}, which is a singlet thus supposed to develop the vacuum expectation value any temperature. This vacuum expectation value should be very small, otherwise it would contribute significantly to the pressure of the system, contradicting with the lattice data. This argument is applicable to any other quarkless-baryon loops.

\subsubsection{Lattice fitting of the potential parameters in the 4-6 PLM}

The potential parameters are constrained as follows.  
First, the trace of PL field ($l$) needs to asymptotically get close to $1$ when one approaches the classical limit 
where $T \to \infty$. 
This condition can be realized by taking 
\begin{equation}
\frac{\partial V(l,T\rightarrow\infty)}{\partial l} \Bigg|_{l=1}=0, 
\label{classical-limit}
\end{equation} 
or assuming the SB limit value, as done in Eq.(\ref{a0}). 
The second condition comes from the FOPT nature, as in Eq.(\ref{GEQ}):    
\begin{equation} \label{second-poly}
\begin{aligned}
&\mathcal{U}(l_0=0,T_c)=\mathcal{U}(l_c,T_c)
\,, \\
&\frac{\partial \mathcal{U}(l,T_c)}{\partial l} \Bigg|_{l=l_0=0}
=\frac{\partial \mathcal{U}(l,T_c)}{\partial l} \Bigg|_{l=l_c}=0 
. 
\end{aligned}
\end{equation}

We have particular interest in the vicinity of the confinement-deconfinement PT at $T=T_c$ and consistency with the large $N$ scaling 
for the thermodynamic quantities around there.  
In that case, it turns out that 
the parameters $a_1$ and $a_2$ are fairly insensitive to this FOPT criticality: 
Near the critical temperature $T_c$, the potential can be simplified as $V=\frac{1}{2} \sum_{i=0}^3 a_i |l|^2+b|l|^4+c|l|^6$, where among the parameters $a_i$, the $a_3$ term is most dominant because of its lowest power contribution by $T_c T$ around $T=T_c$ in the potential. 
Hence the potential 
around $T=T_c$ should be almost controlled by 
the $a_3$ term, so should the pressure. 
Taking into account the classical limit satisfied by  Eq.(\ref{classical-limit}), we see that the $a_0$ term is also relevant.

\begin{figure}[htbp] 
\centering 
\includegraphics[width=0.45\textwidth]{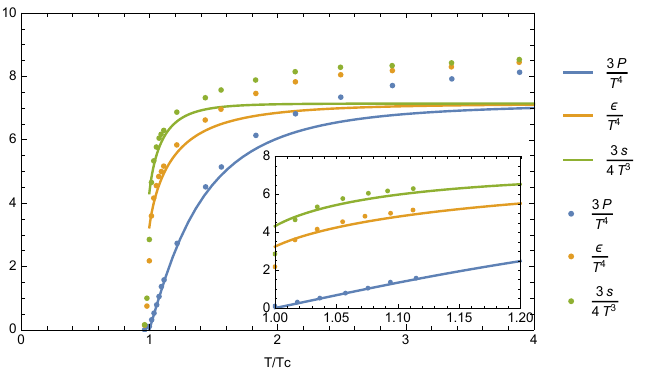} 
\includegraphics[width=0.45\textwidth]{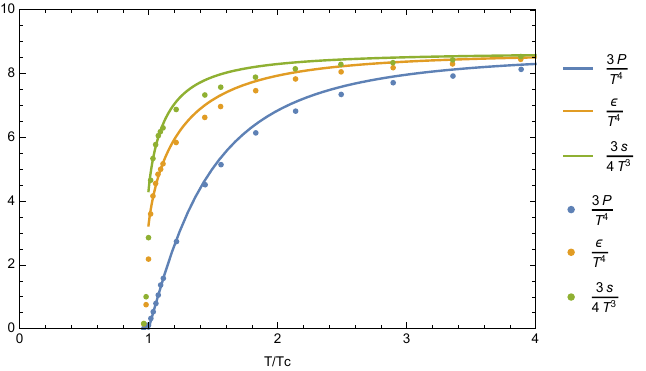} 
\caption{Fitting the 4-6 PLM to the thermodynamic 
quantities of $SU(4)$ PYM theory observed in the lattice simulations~\cite{Datta:2010sq}, 
at around the critical temperature ($T_c$) for the confinement-deconfinement phase transition of the first order.  
In the left panel, several thermodynamic quantities are plotted 
as a function of $T/T_c$ 
with the best fit parameters 
$a_0=4.95,a_1=0,\ a_2=0,\ a_3=-6,\ b=-2.19$ and $c=2.28$, 
while those with $a_0=6.36,\ a_1=0.04,\ a_2=-2.68,\ a_3=-4.72,\ b=-2.28$ and $c=2.58$ are displayed in 
the right panel. 
The LS fit has been worked out only for $1\le T/T_c \le 1.5$ (focusing 
on the criticality regime) in the left panel, as zoomed-in the 
close-up small window, 
while in the right panel for $1 \le T/T_c \le 4$.} \label{fitting-plots}
\end{figure}

The relevant parameter space is then constrained by the conditions in Eqs.(\ref{classical-limit}) and 
(\ref{second-poly}), together with the desired large $N$ scaling of the latent heat in Eq.(\ref{Lat}) along with Eq.(\ref{LN-def}). 
In the case with $a_1 = a_2=0$, those constraints read 
\begin{equation}\label{a3equal}
\begin{aligned}
&0=-a_0+4b+6c
, \\
&|l_c|^2=\frac{-b+\sqrt{b^2+4(a_0+a_3)c}}{2c} 
, \\
&0=-(a_0+a_3)+4b |l_c|^2 
+6 c |l_c|^4 
, \\
& 
(N^2-1)\left( 0.388-\frac{1.61}{N^2} \right) 
\simeq - \frac{3a_3}{2} |l_c|^2 
. 
\end{aligned}
\end{equation}
Thus, all the potential parameters can scale with $(N^2-1)$ in the large $N$ limit, 
consistently with the lattice observation. 
Figure~\ref{fitting-plots} shows that 
the 4-6 PLM fits to the lattice $SU(N)$ thermodynamics with data from~\cite{Datta:2010sq}, 
with focusing on the confinement-deconfinement FOPT 
at around $T=T_c$. 
The left panel has been drawn by assuming $a_1=a_2=0$, while 
the right panel made with assuming nonzero $a_2$. 
Comparison of both panels indeed implies the insensitivity of $a_1$ and $a_2$ 
to the thermodynamics around the criticality. 
See also Table~\ref{table-1}, which more precisely shows 
the best fit 
to the lattice data points 
for the $N=4$ 4-6 PLM without those parameters.

Of importance is to note that the 4-6 PLM with $N=4$, which has been fixed  by fitting to the lattice data as above, can be used to  
construct another 4-6 PLM with $N \ge 5$:  
Given certain $a_3$ for some $SU(N)$ PLM, 
using the large $N$ scaling 
$a_3\rightarrow\frac{M^2-1}{N^2-1}a_3$ from $N$ to $M$ 
and Eq.~(\ref{a3equal}), 
one can immediately get a 4-6 PLM for $SU(M)$.  
This is possible because the last equation in Eq.~(\ref{a3equal}) tells us that, in the large $N$ limit, the parameter $a_3$ is approximately proportional to $N^2-1$ and other equations do not contain direct $N$ dependence. 
Note also that the lattice data tell us thermodynamic quantities $\propto N^2-1$ and the large $ N$ scaling of the latent heat $ \approx N^2-1$. 
Therefore, if $N$ is large enough then all relevant parameters for the $SU(M)$ 4-6 PLM should scale like  
\begin{equation}
a_0\rightarrow\frac{M^2-1}{N^2-1}a_0,\ a_3\rightarrow\frac{M^2-1}{N^2-1}a_3,\ b\rightarrow\frac{M^2-1}{N^2-1}b,\ c\rightarrow\frac{M^2-1}{N^2-1}c
\,, \label{V-scaling}
\end{equation}  
and similar scaling laws for $a_1$ and $a_2$. 

 Note here that in performing the fitting procedure,  
 quoting the available lattice results has been 
 restricted to those 
 which exhibit both the thermodynamics quantities and the latent heat on the same lattice setting. 
 Otherwise, we will create a ``prospected" data from 
 the available ones, by the large $N$ scaling. 
 Currently, only the $N=3,4,6$ lattice simulations have so far been worked out for both those two~\cite{Datta:2010sq}~\footnote{
 The literature~\cite{Panero:2009tv} has actually provided 
 the data on the thermodynamics quantities for $N=5, 6, 8$, 
 though not measuring the latent heat. 
 We have checked that our large $N$-fitting procedure can 
 almost completely reproduce 
 those ``undesired" data (around the criticality).  
 }. 
 Since the reported errors on the data for the thermodynamics quantities are 
 quite small~\cite{Datta:2010sq}, we work on the least square (LS) fit, and evaluate 
 the size of the systematic error by the root-mean square-deviation (RMSD).

\begin{figure}[htbp] 
\centering 
\includegraphics[width=0.45\textwidth]{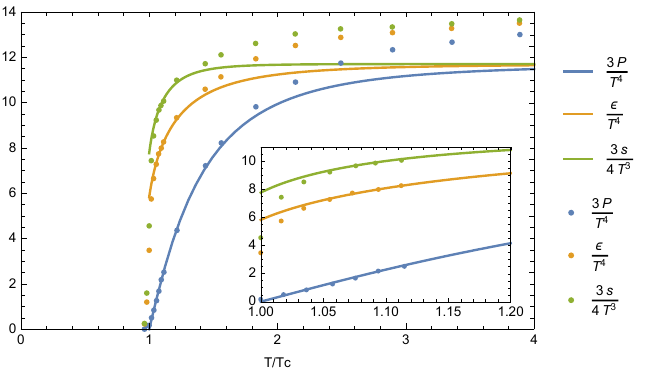} 
\includegraphics[width=0.45\textwidth]{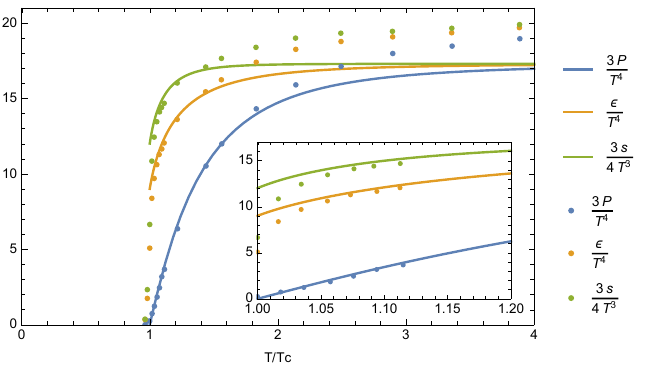} 
\caption{Check on the large $N$ scaling for the 4-6 PLM. 
Left panel: Fitting the 4-6 PLM for $SU(5)$ to the thermodynamic quantities ``measured" in the lattice simulation, where 
the lattice data for $N=5$ have been read off by using the large $N$ scaling for the $N=6$ data in~\cite{Datta:2010sq}, as 
noted in the text.   
Right panel: The same for $SU(6)$. 
For both $N=5$ and $N=6$ cases, we have selected $a_3$ as the 
scaling parameter (scaled from the $N=4$ case), and the other parameters 
have been fixed by the criticality and the latent heat constraints given in Eq.(\ref{a3equal}).   
See Table~\ref{table-1} (4- 6 PLM with $N=5$ and $N=6_1$) 
for the values of the scaled $a_3$ and the other fixed parameters for 
$SU(5)$ and $SU(6)$ cases, and the goodness-of-fit. 
In both panels the LS fits have been performed 
for $1 \le T/T_c \le 1.5$ focusing only around the criticality, as zoomed-in 
the close-up windows. 
} \label{scalingN}
\end{figure}

Figure~\ref{scalingN} and Table~\ref{table-1} 
show how the large $N$ scaling law works  
for $SU(5)$ and $SU(6)$ in Fig.~\ref{fitting-plots} 
(and also the same Table~\ref{table-1}), 
where the desired data are available only for $N=6$, from~\cite{Datta:2010sq}.

Thus the large $N$ scaling procedure allows us to save the fitting steps. 
The procedure will work better if the starting color number is 
set to bigger:
Let $a_{2,3}^\textrm{from $N=m$}|_{N=n} = \frac{n^2-1}{m^2-1} \cdot a_{2,3} |_{N=m}$ be 
the naively scaled $a_{2,3}$ at $N=n$ from $a_{2,3}$ at $N=m$. 
These scaled parameters are 
compared to $a_{2,3}|_{N=n}$ at $N=n$ determined fully by the fitting procedure, 
as follows:  
\begin{equation}
\begin{split}
&\frac{a_2^\textrm{from $N=4$} |_{N=8}}{a_2|_{N=8}} \simeq 0.85\,\,, 
\frac{a_2^\textrm{from $N=6$} |_{N=8}}{a_2|_{N=8}}\simeq 1.11\,\,, 
\frac{a_3^\textrm{from $N=4$} |_{N=8}}{a_3|_{N=8}}\simeq 1.29\,\,,  
\frac{a_3^\textrm{from $N=6$} |_{N=8}}{a_3|_{N=8}}\simeq 0.96 
\end{split}
\,. 
\end{equation}
One can thus see that as $N$ gets larger, the naively rescaled values get closer to 
the ones determined by the fitting procedure.


\subsubsection{Lattice fitting of the potential parameters in the 4-8 PLM}

\begin{figure}[htbp] 
\centering 
\includegraphics[width=0.45\textwidth]{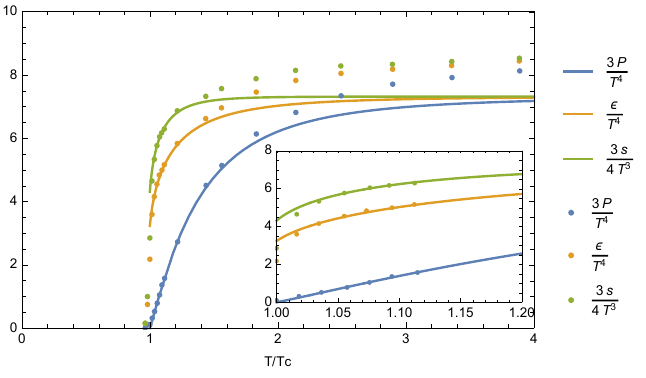} 
\includegraphics[width=0.45\textwidth]{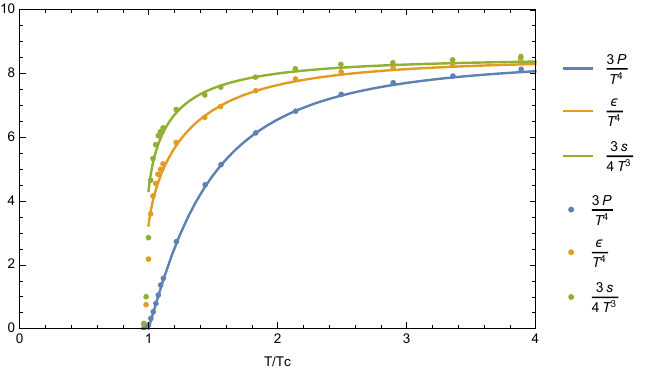} 
\caption{Fitting the $N=4$ 4-8 PLM to the thermodynamic 
quantities of $SU(4)$ PYM theory.  
Left panel:  Thermodynamic quantities as a function of $T/T_c$ 
with the best fit parameters listed in Table~\ref{table-1} 
corresponding to the case $N=4_1$ for the 4-8 PLM. 
Right panel: The same for the case $N=4_2$ with the best-fit parameters in the list of the 4-8 PLM, in Table~\ref{table-1}. 
The LS fit has been worked out only for $1\le T/T_c \le 1.5$ (focusing 
on the criticality regime) in the left panel, as zoomed-in the sub-small window, 
while in the right panel for $1 \le T/T_c \le 4$.
} \label{4-8:PLM}
\end{figure}

Following the same method as done in discussing the 4-6 PLM above, we can analyze the 4-8 PLM. The potential parameter relation then takes the form similar to Eq.~(\ref{a3equal}):  
\begin{equation}\label{8:equal}
\begin{aligned}
&0=-a_0+4b+8c
, \\
&|l_c|^2=-\frac{2\left(\frac{2}{3}\right)^{1 / 3} b}{2\left.(9(a_0+a_3) c^{2}+\sqrt{3} \sqrt{32 b^{3} c^{3}+27(a_0+a_3)^{2} c^{4}}\right)^{1 / 3}}
\\
&\ \ \ \ \ \ \ \ +\frac{\left(9(a_0+a_3) c^{2}+\sqrt{3} \sqrt{32 b^{3} c^{3}+27(a_0+a_3)^{2} c^{4}}\right)^{1 / 3}}{ 2^{4/ 3} \times 3^{2 / 3} c}
, \\
&0=-(a_0+a_3)+b |l_c|^2 
+ c |l_c|^6 
, \\
& 
(N^2-1)\left( 0.388-\frac{1.61}{N^2} \right) 
\simeq - \frac{3a_3}{2} |l_c|^2 
. 
\end{aligned}
\end{equation} 
Thus the 4-8 PLM can also realize the desired large $N$ scaling 
for the thermodynamic quantities in $SU(N)$ PYM theory consistently 
with the large $N$ property of the latent heat. 
The result for fitting to the lattice data is shown in Fig.~\ref{4-8:PLM}, and the best-fit parameters are listed in Table~\ref{table-1}.

Similarly to the 4-6 PLM, this 4-8 PLM can also create a set of  
simple large $N$ copies because of the manifest large $N$ scaling: staring with the 4-8 PLM model having color $N$ 
with the potential parameters fixed, 
one can easily get the corresponding set of the potential parameters for another model with color $M$, just 
by accessing the simple large $N$ scaling for $a_3$ and fitting to the (rescaled) lattice data, to determine the other potential parameters. 

\begin{table} 
\begin{tabular}
{ |p{2.5cm}||p{1.5cm}|p{1.5cm}|p{1.5cm}|p{1.5cm}|p{1.5cm}|p{1.5cm}|p{1.5cm}|p{1.5cm}|   } 
 \hline
 \multicolumn{9}{|c|}{Parameter for 4-6 PLM} \\
 \hline
Color number& $a_0$ & $a_1$ &$a_2$&$a_3$&$b$&$c$&RMSD$^2$ &RE\\
 \hline
${N=4}_1$& 4.9522 &0& 0&-6&-2.1877&2.2839&0.166&0.011\\
${N=4}_2$& 6.3599&0.0371&-2.6796&-4.7265&-2.2827&2.5818&0.320&0.021\\
$N=5$& 7.1365& 0&0&-9.6&-4.5676&4.2345&0.085 &0.004\\
$N=6_1$&9.5886  & 0      &  0       &-14.0 &-7.7105&6.7385& 0.438&0.013\\
$N=6_2$&10.1659 & -0.0629&  -7.3711&-8.5296&-9.7462&8.1918&1.112 &0.032\\
$N=7$&12.3551& 0&  0&-19.2&-11.5643&9.7687& 1.385&0.029\\
$N=8_1$&16.2160& 0&  0&-25.2&-15.1781&12.8214& 2.224&0.035\\
$N=8_2$&14.8974& -0.1011& -11.9253&-15.9985&-20.6588&16.2555&2.244 &0.037\\
 \hline
 \multicolumn{9}{|c|}{Parameter for 4-8 PLM} \\
 \hline
Color number& $a_0$ & $a_1$ &$a_2$&$a_3$&$b$&$c$&RMSD$^2$&RE\\
 \hline
${N=4}_1$& 5.54 & 0&   0  &-6    &-0.7261&1.06  &0.143& 0.010\\
${N=4}_2$& 6.24 & 0&-3.70 &-3.2  &-0.9794&1.27  &0.103& 0.007\\
$N=5$    & 9.22 & 0&-5.75 &-5.19 &-2.25  &2.28  &0.067& 0.003\\
$N=6$    & 12.02& 0&-8.42 &-7.18 &-4.29  &3.65  &0.183& 0.005\\
$N=7$    & 14.70& 0&-11.71&-9.32 &-7.15  &5.41  &0.475& 0.010\\
$N=8$    & 16.73& 0&-16.29&-10.98&-11.33 &7.76  &0.912& 0.014\\
$N=9$    & 20.20& 0&-20.69&-13.94&-15.29 &10.17 &1.810& 0.023\\
$N=10$   & 24.00& 0&-25.60&-17.25&-19.78 &12.89 &3.222& 0.033\\
$N=11$   & 28.14& 0&-31.02&-20.91&-24.77 &15.90 &5.288& 0.044\\
 \hline
\end{tabular}
\caption{The fitting parameters for the 4-6 and 4-8 PLMs. 
The two groups on the list for the cases with $N=4,6,8$ separate two 
categories: one is the case where only $a_3$ is used as the fitting parameter (corresponding to $N_1$), and the other case treats all $a_1,a_2$, and $a_3$ as the 
fitting parameters ($N_2$). 
The $n$ denotes the number of data and the LS fits for the $N=4$ PLMs have been performed based on the data for $T \leq 1.2T_c$. 
As to the cases with $N=5$ and $N\ge 7$, which are directly unavailable from~\cite{Datta:2010sq}, 
the prospected data have been read off by using 
the large $N$ scaling from $N=4$ or $6$ data, as noted in the text, 
and the fitting has been done by using the method described in the text. 
Though their RMSDs ($=$LS$/\sqrt{n}$) are relatively poorer,  
the relative errors (RE), defined by RMSD$^2/(N^2-1)$, remain small. 
This implies that 
the fitting still works well.
}  
\label{table-1}
\end{table}

\section{GW spectra based on the 4-6 and 4-8 PLMs} \label{GW}

\subsection{Dark confinement-deconfinement FOPT in a secluded gluonic plasma}

The GW associated with the confinement-deconfinement 
PT would be generated from the bubble nucleation of the confinement vacuum.  The bubble nucleation rate per spacetime, $\Gamma$, contains two contributions.  
The first one comes from its thermal fluctuation, $\Gamma(T)$, 
and the second one generated by the quantum effects. 
In the present study, we ignore the second contribution because 
the quantum effects would dominate only in supercooling.

The thermal contribution to the bubble nucleation rate per volume per time is generically 
given by
\begin{equation}
\Gamma(T)\approx A T^4 e^{-\frac{S_3(T)}{T}}
, \label{GammaT}
\end{equation}
where $A$ is supposed to be of ${\cal O}(1)$, and 
$S_3(T)$ is the $O(3)$ symmetric action in three-dimensional Euclidean space: 
\begin{equation}
S_3(T)=4\pi \int_0^{\infty} R^2dR \left[\frac{1}{2}\L\frac{d\phi}{dR}\R^2+V(\phi,T)\right],
\end{equation}
with $R=|\vec{x}|$. 
$V(\phi,T)$ denotes an effective potential for two vacua (false vacuum and true vacuum), and 
$\phi(R)$ is the bounce solution satisfying the Euclidean equation of motion
\begin{equation} \label{EOM}
\begin{split}
\f{d^2\phi}{dR^2}+\f{2}{R}\f{d\phi}{dR}=V',
\end{split}
\end{equation}
with $V' \equiv \partial V(\phi, T)/\partial \phi$, and the boundary conditions $\underset{R\ra \infty}{\lim}{ \phi(R)}=0$ (at the false vacuum position) and $\f{d \phi(R)}{dR}|_{R\ra 0}=0$. 

One thing to note is that the generic bounce solution has mass dimension 1, so that the exponent $S_3(T)/T$ in Eq.(\ref{GammaT}) is dimensionless. 
On the contrary, in the PYM case, the role of the bounce solution is played by the 
trace of the PL field, $l$, which is dimensionless.   
To match the case with the generic thermal evolution in Eq.(\ref{GammaT}),  
we need to rescale $l$ so as to give mass dimension 1 to $l$, 
like $l \to  l  \Lambda$, and fix the scale parameter as $\Lambda = T$. 
This scaling would be reasonable because $l$, or $L$ in Eq.(\ref{PL:def}),  
involves only 
the temperature $T$ as the dimensionful parameter. Similar prescription has been applied in the literature~\cite{Pisarski:2000eq}.

 Another thing to note is the distinction between the temperature in the dark gluonic bath and that in the bath made of the standard model (SM) particles. We label them respectively as $T$ (for the dark PYM sector) and ${\cal T}$ (for the SM sector). 
Since PYM theory incorporates no matters, the two baths should decouple and thermalize individually. In $\Gamma(T)$, Eq.(\ref{GammaT}), 
the prefactor $T^4$ originates from the spacetime volume evolved by 
the scale factor $a$ of the universe as $V=a^3t$, so 
should be identified simply with neither ${\cal T}$ nor $T$. 
However, the difference in $T$ and ${\cal T}$ actually 
does not matter so much since it merely gives a logarithmic correction to the nucleation temperature $T_n$. 
Whereas both temperatures in $S_3(T)/T$ (the argument of $S_3$ and 
the normalization factor for the exponential, $1/T$) should be associated with the dark PYM-sector: In $S_3(T)$, $T$ is supplied from a given PLM, and the origin of $1/T$ is traced back to the imaginary-time formalism in the thermal field theory, where $T$ should be the temperature of the thermal bath 
created in the thermalized dark PYM.

In the temperature region of interest (say, before the big-bang nucleosynthesis (BBN)) where both baths are in the form of radiation~\footnote{This is imprecise for the dark gluonic sector when it enters the sQGP region and further approaches the transition, where the equation of matter deviates from radiation more significantly. But our discussions approximately hold as long as the universe is not dominated by the PYM sector.}, then $T_i=\L\f{30\rho_i}{\pi^2}g_i^*\R^{1/4}$ with $T_1=T$ and $T_2={\cal T}$, and similar for others.  Here 
$\rho_i$ and $g_i^*$ are the energy density and relativistic degrees of freedom in the bath, respectively.  
Further assuming that the entropy is conserved, then both temperatures scale as $1/a$, and are expanding with the Hubble parameter 
\begin{equation} \label{}
\begin{split}
H=\sqrt{\L\rho_{\rm SM}+\rho_{\rm PYM}\R/3M_{\rm Pl}^2}.
\end{split}
\end{equation}
As a consequence, in the thermal epoch under consideration, the ratio $\zeta\equiv {\cal T}/T$ keeps a constant, determined by the production mechanism of the two sectors. To the end of having a correct relic density of dark glueball dark matter, $\zeta\gg 1$ is favored~\cite{Kang:2019izi}.  But it will significantly suppress the amplitude of GW. 
Thus in this article we just take $\zeta$ to be merely a parameter, and  
suppose a situation that the dark glueball is not the main component of dark matter, or not the dark matter candidate at all. 
Such a setup justifies our treatment later, when the predicted GWs are discussed 
in comparison between the case with $\zeta =0$ and $\zeta=1$ (Figs.~\ref{SU(4)T} and~\ref{SM}). 
Anyway, in what follows we will use $T$ only, and the SM-sector temperature is simply $\zeta T$.

We use the Mathematica program AnyBubble~\cite{Masoumi:2017trx} to numerically solve Eq.~(\ref{EOM}), and evaluate $S_3(T)/T$ and the bubble nucleation rate $\Gamma(T)$ in Eq.(\ref{GammaT}). 
The criterion for bubble nucleation is that at the nucleation temperature $T_n$, a single bubble is nucleated within one Hubble horizon volume; it roughly amounts to the condition  $\Gamma(T_n)\sim H(T_n)$. In the radiation dominated universe, this condition reads 
$S_3( T_n)/T_n\sim140$, and see Ref.~\cite{Kang:2020jeg} for more  detailed discussions. 
Through this equation we determine the bubble nucleation temperature $T_n$. 

With the bounce action and $T_n$ at hand, we can calculate  two key parameters characterizing the FOPT, 
which almost completely determine the GW spectra. 
One is the  $\alpha$ parameter, which measures the strength of the FOPT via the ratio between the latent heat and the total energy density of the universe at $T=T_n$, 
\begin{equation} 
\begin{aligned}
&\alpha\equiv \frac{L_N(T_n)}{\rho_{\rm SM}(\zeta T_n)+\rho_{\rm PYM}(T_n)},
\label{alpha}
\end{aligned}
\end{equation}
The other one is $\wt \beta$, the inverse time scale of PT duration in terms of the Hubble time scale, 
\begin{equation} 
\begin{aligned}
\wt \beta\equiv -\f{1}{H}\f{d}{dt} \left(\frac{S_3(T)}{T}\right)
\Bigg|_{t=t_{n}}, 
\end{aligned}
\end{equation}
with $t_n$ being the cosmic time corresponding to ${T}_n$. It is convenient to replace $t$ with $T$, via the relation  $\frac{dt}{dT}=\frac{-1}{H T}$, and then $\wt \beta$ goes like 
\begin{equation} 
\begin{aligned}
\wt \beta=T_{n} \frac{d}{dT}\left(\frac{S_3(T)}{ T}\right)
\Bigg|_{T=T_{n}} 
\label{betat}
\end{aligned}
\end{equation}
Now, both $\alpha$ and $\beta$ are calculated as a function of the dark - PYM sector temperature $T$ (given certain $\zeta$). 
The resultant numbers for the parameters relevant to the GW spectra are give in Table~\ref{table-2}. 
We see that the typical $\wt \beta$ is as huge as $10^5$, namely a very fast PT. 
That short duration of PT suppresses the production of GW. 
This result turns out to be 
fairly irrespective to whether $\zeta$ is zero or nonzero (See Figs.5 and 7, and related discussions around there).  

\begin{table}
\begin{tabularx}{0.9\textwidth} { 
  || >{\centering\arraybackslash}X 
  | >{\centering\arraybackslash}X 
  | >{\centering\arraybackslash}X 
  | >{\centering\arraybackslash}X 
  | >{\centering\arraybackslash}X 
  | >{\centering\arraybackslash}X 
  | >{\centering\arraybackslash}X 
  | >{\centering\arraybackslash}X | }
 \hline
Model and $N$ & 4-6 PLM ${N=4}_1$& 4-6 PLM ${N=4}_2$& 4-6 PLM $N=5$& 4-6 PLM $N=6$ & 4-6 PLM $N=7$& 4-6 PLM $N=8$  & 4-8 PLM ${N=4}_1$\\
 \hline
$T_c$/$g_*$ & 100\, ${\rm GeV}$  /133  & 100\, ${\rm GeV}$  /133  & 100\, ${\rm GeV}$  /142 & 100\, ${\rm GeV}$  /153& 100\, ${\rm GeV}$  /166 & 100\, ${\rm GeV}$  /181 & 100\, ${\rm GeV}$  /133 \\
\hline
$T_n$ & 0.9968$T_c$  & 0.9973$T_c$  & 0.9960$T_c$ & 0.9956$T_c$& 0.9954$T_c$ &0.9957$T_c$ & 0.9981$T_c$  \\
\hline
$\alpha$ & 0.094  & 0.095 & 0.160 & 0.229& 0.300  & 0.363& 0.095   \\
\hline
$\wt \beta$  & 97089 & 111073 & 75306 & 67404& 64897 &68694 & 159536   \\
\hline \hline 
4-8 PLM ${N=4}_2$ & 4-8 PLM $N=5$ & 4-8 PLM ${N=6}$& 4-8 PLM ${N=7}$& 4-8 PLM $N=8$& 4-8 PLM $N=9$& 4-8 PLM ${N=10}$& 4-8 PLM ${N=11}$ \\
 \hline
100\, ${\rm GeV}$  /133 & 100\, ${\rm GeV}$  /142  & 100\, ${\rm GeV}$  /153  & 100\, ${\rm GeV}$  /166  & 100\, ${\rm GeV}$  /181 & 100\, ${\rm GeV}$  /198& 100\, ${\rm GeV}$  /217 & 217\, ${\rm GeV}$  /238 \\
\hline
0.9973$T_c$ & 0.9962$T_c$& 0.9951$T_c$  & 0.9942$T_c$  & 0.9932$T_c$ & 0.9931$T_c$& 0.9932$T_c$ & 0.9933$T_c$ \\
\hline
 0.095  & 0.160& 0.229  & 0.299 & 0.367 & 0.394 & 0.494 & 0.550  \\
\hline
 111497 & 80040 & 61638 & 51869 & 43684 & 43204 & 43240 & 43657  \\
\hline
\end{tabularx}
\caption{The reference values for the parameters relevant to 
the GW spectra, predicted from the 4-6 and 4-8 PLMs with $N=4-8$, and $N=4-11$, 
respectively. 
The subscript attached on 4 of $N=4$ denotes the same category 
as introduced in Table~\ref{table-1}. 
The SM contribution parameter $\zeta$ has been set to 1. 
}
\label{table-2}
\end{table}

Let us end up with this subsection by stressing an interesting observation of $\wt \beta$. That is, it shows a minimum at some color number $N_m$. This feature has been pointed out in Refs.~\cite{Huang:2020mso,Halverson:2020xpg}, but we would like to stress that the value of $N_m$ depends on the model. For the model in Ref. \cite{Huang:2020mso},  $N_m=6$, while  $N_m=4$ in Ref.~\cite{Halverson:2020xpg} which adopts a very different model, the matrix model.  In our model, we find that $N_m=7 (9)$ for the $4-6 (8)$ PLM; see Table.~\ref{table-2}. 

The above feature may be partially understood. We can prove that  $\wt \beta$ decreases with $N$ by the large $N$ scaling. For a sufficiently large $N$, Eq.~(\ref{EOM}) can be solved once and for all, due to the scaling behavior of the effective potential $V_N=N^2U$ with $U$ independent of $N$. We assume that  temperature is fixed, hence shall drop its dependence. 
Writing the bounce solution for the potential $U$ as $\phi(R)$, we see that 
the bounce solution for $V_N$ is simply given by $\phi_N(R)=\phi(NR)$. 
With this solution, the three-dimensional Euclidean action $S_3$ for color $N$ 
can be rewritten in the following form: 
\begin{equation}
    S_3=\frac{2\pi}{N}\int^{\infty}_0 r^2dr\left[\frac{1}{2}\left(\frac{d\phi}{dr}\right)^2+U\right]. 
\end{equation}
Now it is evident that $S_3$ simply scales as $1/N$. 
Note that our $T_n$ is always very close to $T_c$, which is determined by $U$  universal for the large $N$. 
Thus the nucleation condition $S_3/T_n\simeq 140$  leads to a slightly larger $T_n$ when increasing $N$. Moreover, the slope of $S_3/T$ sharply increases when the temperature approaches $T_c$. Therefore, $\wt\beta$, which is  proportional to  the slope, increases with $N$. 
Taking into account the fact (not proved, just a numerical observation) that $\wt \beta$  decreases with $N$ for a small $N$, we deduce that $\wt \beta$ reaches the minimum  at some moderately large $N$.

\subsection{GW sources from the fast PT}

In the cosmic FOPT, there are three kinds of sources for GWs, i.e., bubble collision, sound wave and magnetohydrodynamic (MHD) turbulence. The latent heat released during the FOPT distributes into these sources, but the fractions are not precisely determined yet. Usually, it is believed that the fraction of the latent heat transferred to the bubble collision is negligible, $\kappa_{col}\ll 1$,  
provided that $\alpha$ does not become extremely large~\cite{Ellis:2019oqb}. The GW sources are dominated by the two bulk motions of the plasma. 

One bulk motion is the sound wave propagating in the plasma after the percolation, and the latent heat that goes into it is estimated to be~\cite{soundwave}
\begin{equation} 
\kappa_{sw}\approx \alpha(0.73+0.083\sqrt{\alpha}+\alpha)^{-1},
\end{equation}
suppressed by the small $\alpha $. The GW peak frequency at $T_n$ is $f_{sw,*}=2(8\pi)^{1/3}/[\sqrt{3}(v_w-c_s)R_*]$ with $R_*$ being the average bubble separation at the collision and the speed of sound in the plasma $c_s=1/\sqrt{3}$. This separation is related to the typical time scale of PT via $R_*=(8\pi)^{1/3}v_w/\beta_n$, given the exponential approximation of $\Gamma(T)$ around $T_n$. The observed GW spectra peak at  $f_{sw}=f_{sw,*}a_0/a(T_n)$, as a result of redshift from the GW production time $t_n$ to today. 
Then the  peak frequency is parameterized as \cite{Ellis:2019oqb}
\begin{equation} \label{} 
\begin{split}
f_{sw}&=1.65\times 10^{-5}\L\frac{T_n}{100 \rm GeV}\R\L\frac{g_*}{100}\R^{\frac{1}{6}} \frac{3.4}{(v_w-c_s)H_* R_*} \rm Hz
\\
&=2.75\times10^{-5}\frac{\wt\beta}{v_{w}}\L\frac{T_n}{100 \rm GeV}\R\L\frac{g_*}{100}\R^{\frac{1}{6}} \rm Hz.
\end{split}
\end{equation}
The amplitude of the GW spectrum for the sound wave is then given by~\cite{Ellis:2019oqb, Caprini:2019egz}
\begin{align} 
h^2\Omega_{sw}(f)&=6.35\times 10^{-6}\L H_* R_*\R \L H_* \tau_{sw} \R  \L\frac{ \kappa_{sw}\alpha}{1+\alpha}\R^{2}\L\frac{100}{g^*}\R^{\frac{1}{3}}v_{w} S_{sw}(f),
\end{align}
where the shape of the spectrum takes the form 
\begin{align} 
S_{sw}(f)&=(f/f_{sw})^3\left[\frac{7}{4+3(f/f_{sw})^2}\right]^{\frac{7}{2}}.
\end{align}
The amplitude is proportional to $\tau_{sw}$, the duration of the sound wave source, defined as $H_*\tau_{sw}=\min[1,H_* R_*/U_f]$ with $U_f$ being the root-mean square -fluid velocity~\cite{simulations}
\begin{equation} 
U_f\simeq \f{\sqrt{3}}{2}\L\f{\alpha(1-\kappa_{col})}{1+\alpha(1-\kappa_{col})}\kappa_{sw}\R^{1/2}.
\end{equation}
A smooth expression for $H_*\tau_{sw}$ has been derived in Ref.~\cite{Guo:2020grp}, and it is simply reduced to $H_* R_*/U_f$ in the limit $H_* R_*/U_f\ll 1$. This is true for almost all confinement-deconfinement PTs, which will generate huge $\beta$. Consequently, the sound wave source is considerably suppressed by such a factor. So, in this case, the GW from the sound wave becomes
\begin{align} 
h^2\Omega_{sw}(f)&=6.3\times 10^{-5}\f{1}{\wt\beta^2}\L\frac{ \kappa_{sw}\alpha}{1+\alpha}\R^{2}\L\frac{100}{g^*}\R^{\frac{1}{3}}v_{w}^2 S_{sw}(f).
\end{align}
Regarding the bubble wall velocity, we first note that 
the background field $ l$ does not interact with the SM particles, and therefore the bubble expansion seems not to be hampered by the SM plasma. 
Even without the SM plasma, the dark gluonic plasma itself would provide friction during the bubble expansion, which should be too nonperturbative to simply be quantified.   
In the present analysis, we will take the bubble wall velocity $v_w=1$ 
to shed light only on the maximized amplitude. 
We will come back to this point later in the discussion section.

When the sound wave period ends and the fluid flow becomes non-linear, then the other fluid bulk motion, the MHD turbulence is generated. If $\tau_{sw}$ is relatively long, at least one Hubble time scale, then the MHD turbulence is suppressed, having an efficiency factor $\kappa_{turb}\sim0.05\kappa_{sw}$~\cite{simulations}. However, as shown above, $\tau_{sw}\ll 1/H^*$ and consequently $\kappa_{turb}$ may be significantly enhanced. But  the quantitative enhancement is unknown yet owing to the fact that a part of the fluid motion not consumed by the sound speed wave can convert into heat. Optimistically, we assume that all of the energy transfers to turbulence, giving rise to the GW spectrum~\cite{Caprini:2009yp},
\begin{equation}\label{TURB}
h^2\Omega_{turb}(f)=3.3\times10^{-4}\f{1}{\wt\beta}\L 1-\f{2(8\pi)^{1/3} v_w}{\sqrt{3}\wt\beta}\R\L\frac{ \kappa_{sw}\alpha}{1+\alpha}\R^\frac{3}{2}\L\frac{100}{g}\R^{\frac{1}{3}}v_{w} S_{turb}(f),
\end{equation} with the shape function
\begin{equation} 
\begin{split}
S_{turb}(f)=\frac{(f/f_{turb})^3}{[1+(f/f_{trub})]^{\frac{11}{3}}(1+8\pi f/h)}.
\end{split}
\end{equation}
which, compared to $S_{sw}(f)$, shows a moderately large suppression $\sim{\cal O}{(10)}$ in the high frequency region. The peak frequency is similar to that of the sound wave source~\cite{Ellis:2019oqb}, 
\begin{equation} \label{}
\begin{split}
f_{turb}&=1.65\times 10^{-5}\L\frac{T_n}{100 \rm GeV}\R\L\frac{g_*}{100}\R^{\frac{1}{6}} \frac{3.9}{(v_w-c_s)H_* R_*} \rm Hz
\\
&=3.15\times10^{-5}\frac{\wt\beta}{v_{w}}\frac{T}{100 \rm GeV}\L\frac{g}{100}\R^{\frac{1}{6}}\rm Hz.
\end{split}
\end{equation}

\subsection{
Predicted GW spectra 
}

\begin{figure}[htbp] 
\centering 
\includegraphics[width=0.7\textwidth]{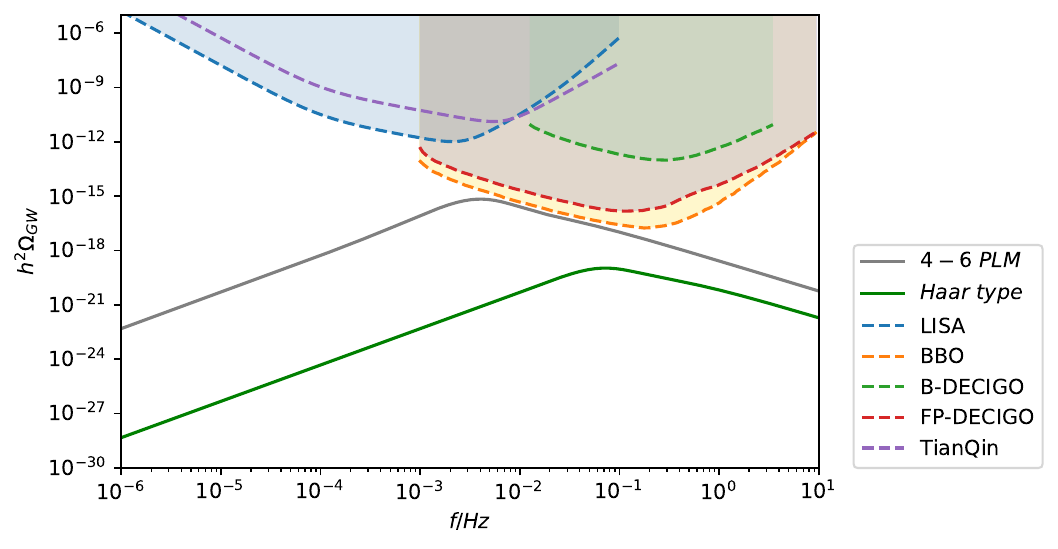} 
\caption{GW spectra predicted from different models for $N=4$
under an extreme scenario that  
the secluded dark gluon plasma to highly dominate over the SM plasma 
(i.e. $\zeta =0$). 
The critical temperature has been set as $T_c=270$ MeV. 
For more details, see the text.} \label{SU(4)} 
\end{figure}

Using the formulae above combined with the 4-6 PLM in the previous section, we plot the GW spectra predicted from the dark $SU(N)$ PYM theory in Fig.~\ref{SU(4)} for $N=4$ 
marked as ``4-6 PLM'', 
in comparison with the prospected GW interferometer sensitivities in the future (LISA~\footnote{We have quoted the $C2$ type of the LISA's configuration in~\cite{Caprini:2015zlo}. 
All GW detection sensitivities used in the present analysis 
have been based on the power-law integrated sensitivity curve.}~\cite{Audley:2017drz,Baker:2019nia}, 
BBO~\cite{Crowder:2005nr,Corbin:2005ny,Harry:2006fi,Thrane:2013oya,Yagi:2011wg}, DECIGO~\cite{Yagi:2011wg,Seto:2001qf,Isoyama:2018rjb} and 
TianQin~\cite{Lu:2019sti}). 
In the figure, as a reference, 
we set $\zeta =0$ (i.e. no SM contributions), 
which can ideally be realized in the large $N$ limit. 
The critical temperature $T_c$ for the confinement-deconfinement PT has been 
set to 270 MeV, just as a reference point inspired by the PYM limit for QCD~\footnote{
In the literature~\cite{Boyd:1996bx} 
$T_c$ was estimated for $N=3$ to be $\sim $ 260 MeV with 
a particular string tension size input.  
}.  
In Fig.~\ref{SU(4)}, we have also made comparison with the Haar-type PLM in Eq.~(\ref{Potential}) (labeled by ``Haar-type''), 
though it is inconsistent with lattice data on thermodynamic quantities in terms of 
the large $N$ scaling as well as the confinement-deconfinement criticality scaling in $T$. 
From Fig.~\ref{SU(4)} we see that the predicted GW signal from the $N=4$ 4-6 PLM 
can have high sensitivity reach to be probed by the prospected BBO detector. 
Note also that the 4-6 PLM's signal is significantly different and gets larger than  
the Haar-type PLM's. 


\begin{figure}[htbp] 
\centering 
\includegraphics[width=0.49\textwidth]{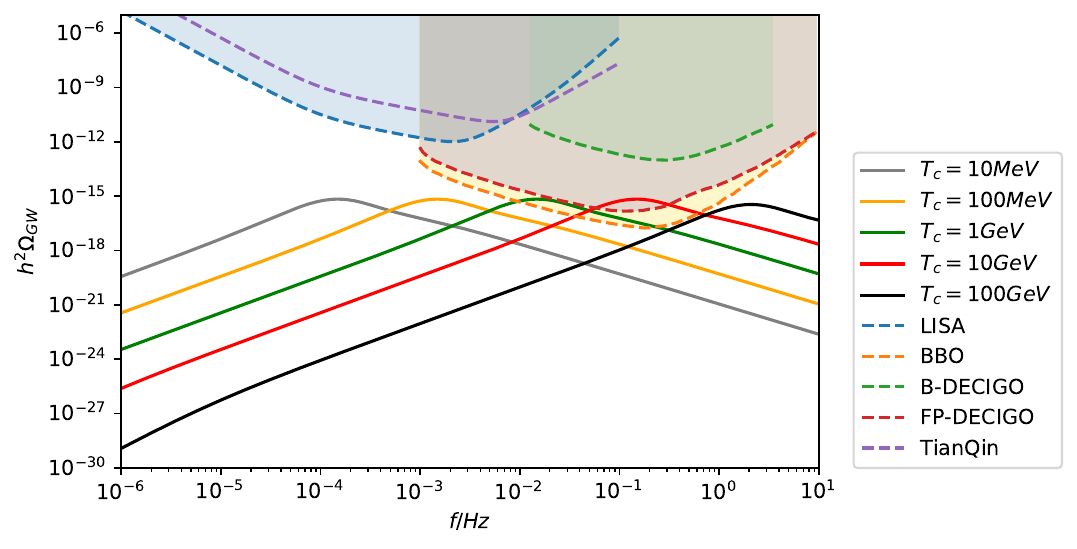} 
\includegraphics[width=0.49\textwidth]{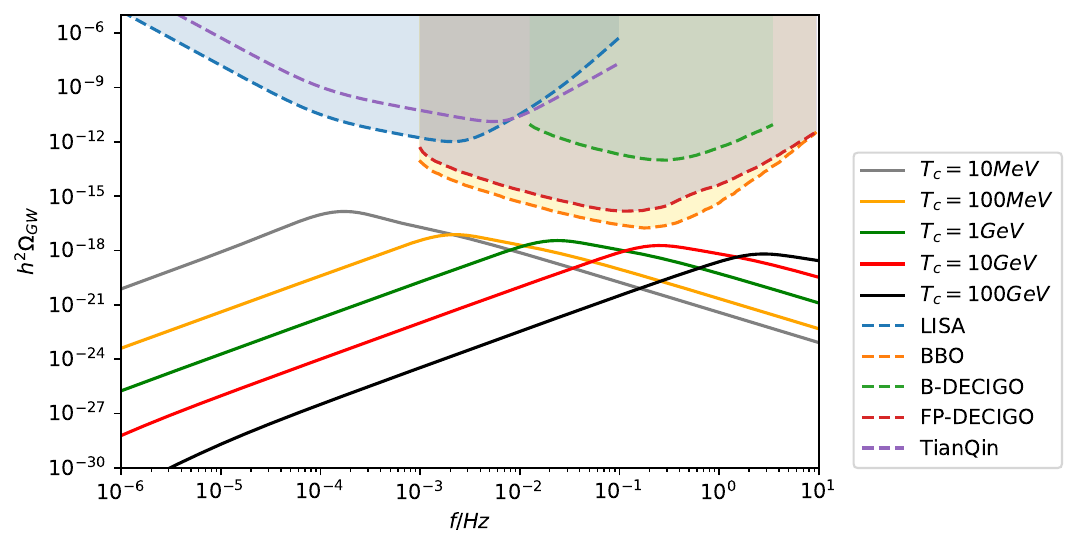} 
\caption{GW spectra with the same condition as in Fig.~\ref{SU(4)}, but 
with varying $T_c$ in several orders. 
The right panel includes the SM contribution by $\zeta=1$, while the 
left panel doe not ($\zeta =0$). 
} \label{SU(4)T}
\end{figure}

\begin{figure}[htbp] 
\centering 
\includegraphics[width=0.49\textwidth]{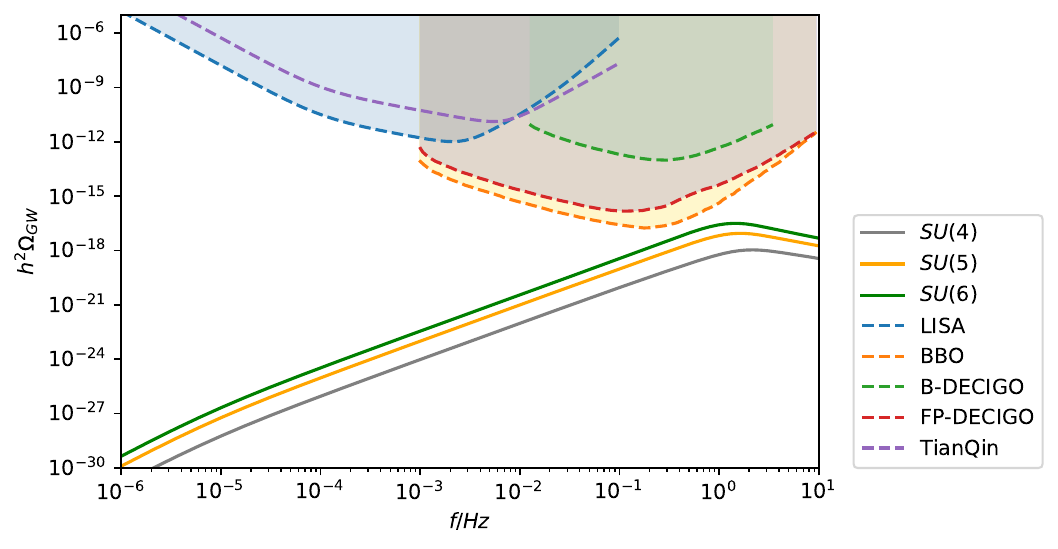} 
\includegraphics[width=0.49\textwidth]{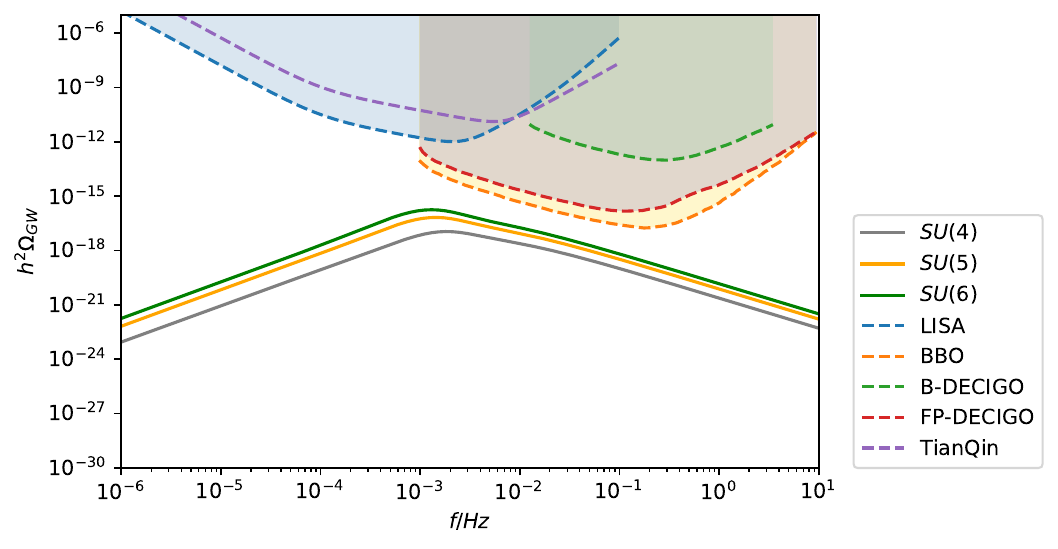} 
\caption{GW signals from the 4-6 PLM  
for $N=4, 5, 6$ with $T_c=100$ GeV (left panel) and $T_c=100$ MeV (right panel). 
The SM contribution has been set as $\zeta=1$. 
The signals with $N =7, 8$ for $4-6$ PLMs 
will look like nearly overlapping with the $N=6$ signal in this plot, 
because of almost degenerate $\tilde{\beta}$s as read off from Table~\ref{table-2}. } \label{SM}
\end{figure}

Varying the input $T_c$ value, we also plot the predicted GW signals 
for $N=4$, 
in Fig.~\ref{SU(4)T}. 
In the left panel, we have taken $\zeta=0$ as in Fig.~\ref{SU(4)}, while 
in the right panel $\zeta =1$. 
As clearly seen from the figure, 
different critical temperatures will only 
shift the peak frequencies of GW and hardly affect the peak amplitudes. 
We also find that the SM contribution give no significant 
contribution to the GW spectra from the dark PYM. 
In Fig.~\ref{SM}, similar plots with $\zeta=1$ 
have been displayed for $N=4,5,6$.

\begin{figure}[htbp] 
\centering 
\includegraphics[width=0.49\textwidth]{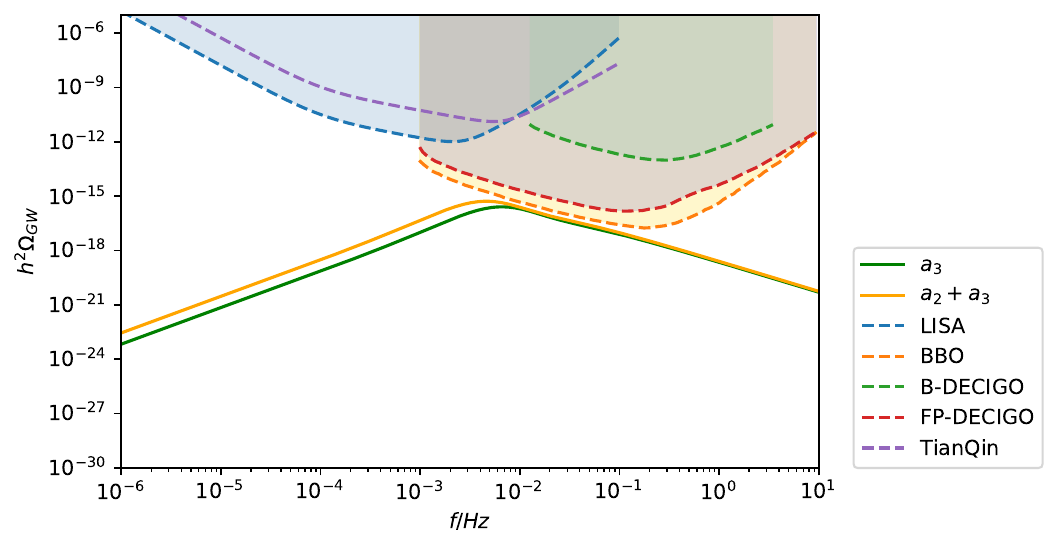} 
\includegraphics[width=0.49\textwidth]{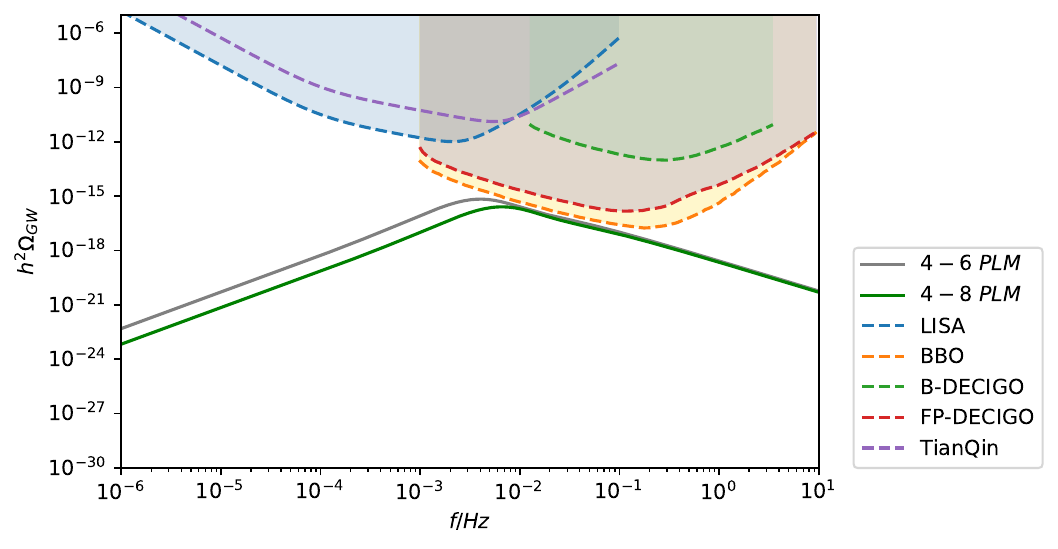} 
\caption{Left panel: GW signals from 4-8 PLMs with $a_2=0$ (labelled by ``$a_3$", corresponding to the category $N=4_1$) and $a_2\neq 0$ (labelled by ``$a_2 + a_3$", corresponding to the category $N=4_2$), 
corresponding to the fitting results in Fig.3, 
for $N=4$ with $T_c=270$ MeV. 
Right panel: Comparison of the GW signals 
from the 4-6 PLM and 4-8 PLM with $a_2=0$, 
for $N=4$ and $T_c=270$ MeV. 
The SM contribution has been turned off by taking $\zeta =0$.} \label{GW:6-8}
\end{figure}

In Fig.~\ref{GW:6-8}  
we show the $N=4$ 4-8 PLM prediction to the GW spectra without the SM term ($\zeta=0$). 
The left panel displays the two scenarios 
with $a_2=0$, i.e. the category $N=4_1$, or with $a_2 \neq 0$, i.e. the category $N=4_2$, while in the right panel, 
comparison between the 4-6 PLM and 4-8 PLM has been shown 
for the case with $a_2=0$, i.e. the category $N=4_1$. 
In the figure we have taken $\zeta=0$ and $T_c$=270 MeV. 
We see that both 4-6 and 4-8 PLMs would be possible to be probed by 
BBO.

\begin{figure}[htbp] 
\centering 
\includegraphics[width=0.49\textwidth]{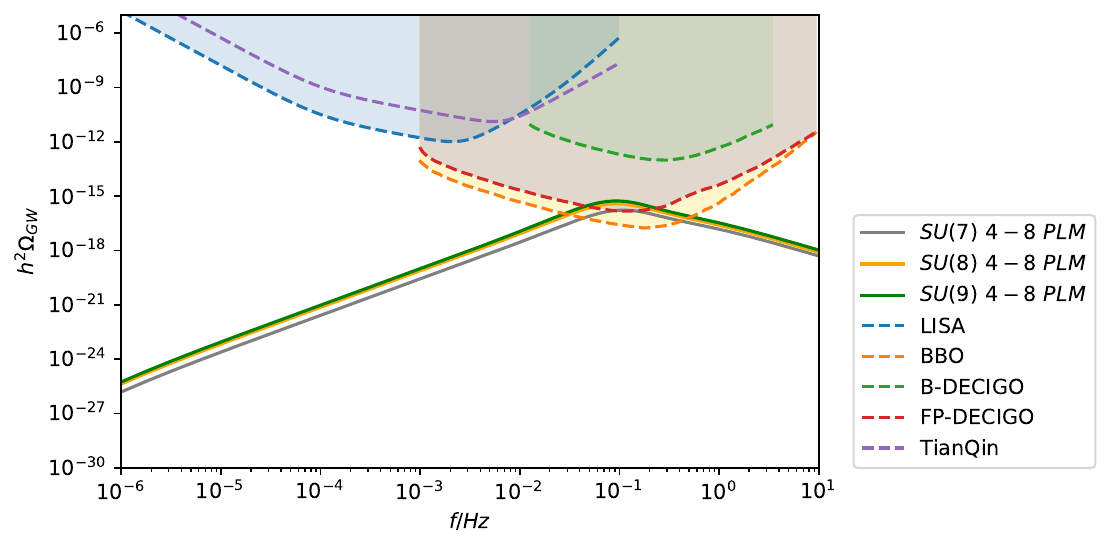} 
\includegraphics[width=0.49\textwidth]{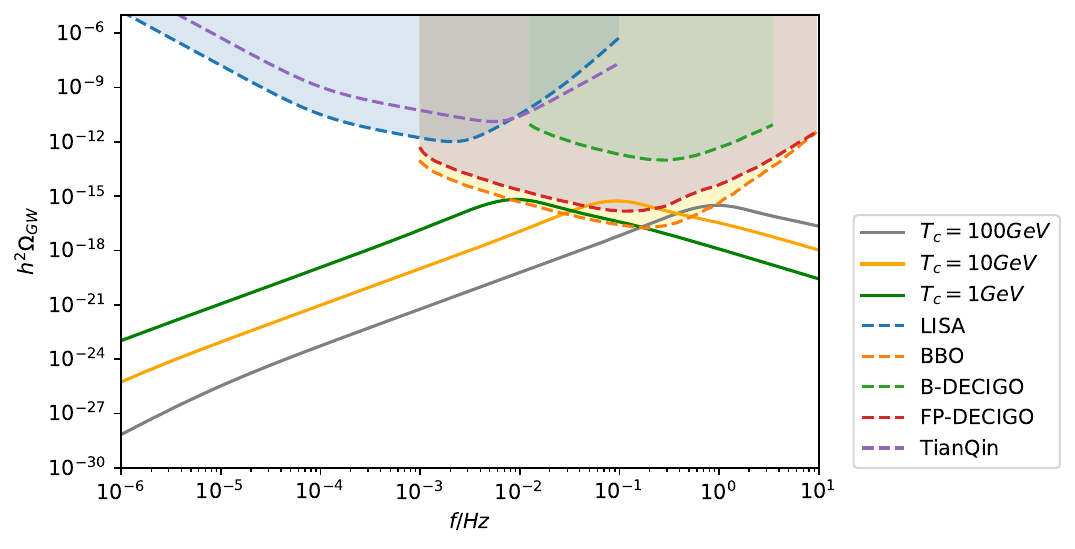} 
\caption{
Left panel: GW signals of $4-8$ PLM PTs with $N=7,8,9$ 
at $T_c=10\gev$. Right panel: 
The dependence of the critical temperature $T_c$ on the GW signals from the 
$N=9$ 4-8 PLM.  
Both panels have taken into account the SM contribution ($\zeta=1$).
} \label{GW:4-8 B}
\end{figure}

Taking larger $N$, the 4-8 PLM sensitivity gets larger and 
reaches the maximum at $N=9$, as expected from the minimum point of 
$\tilde{\beta}$ in Table~\ref{table-2}. 
The left panel of Fig.~\ref{GW:4-8 B} shows the predicted signals for $N=7,8,9$ 
with the SM contribution included ($\zeta=1$) and $T_c=10$ GeV fixed,
and the right panel displays the $T_c$ dependence for the $N=9$ signals.  
The figure implies that the GW spectra from the 4-8 PLM with 
larger $N$ can be probed 
at earliest by BBO and DECIGO.

\section{Conclusion and discussions}

A pure $SU(N)$ Yang-Mills (PYM) sector, for instance, predicted by the string theory, may be an active part of particle physics beyond the standard model (SM). Maybe, such a PYM sector is completely secluded to the SM sector, and then we should peek into this dark sector by means of its gravitational wave (GW) signals, which are generated during the first-order (FO) confinement-deconfinement phase transition (PT). 

The gauge invariant object, the Polyakov loop $L$, associated with the global $Z_N$ symmetry of $SU(N)$ (called the center symmetry), is a well defined order parameter of this PT. And a proper effective model based on $L$, namely the Polyakov loop model (PLM), successfully  describes the confinement-deconfinement PT. In this article we first looked into the widely used Haar-type PLMs, and demonstrated that they fail to produce the large $N$ scaling for the thermodynamic quantities and the latent heat at around the criticality of the PT 
reported from the lattice simulations.  
We then proposed a couple of PLMs with polynomial terms, 
dubbed the 4-6 PLM and 4-8 PLM, with a negative quartic term to trigger the FOPT. 
Such a structure may naturally arise in the presence of a heavy PL with the center charge 2, given that the fundamental or normal PL having charge 1. 
We showed that  those models give the desired thermodynamic  
and large $N$ properties at around the criticality. 
The predicted GW spectra were shown to have high enough sensitivity to be probed in the future prospected interferometers such as  
BBO and DECIGO.

The dark PYM opens novel interesting possibilities in the early universe, and here are some open questions need to be addressed elsewhere:
\begin{itemize}

\item In discussing the bubble velocity, in Sec.III, 
we have argued that the bubble is expanding relativistically. 
However, as was mentioned in the text, 
the dark gluonic plasma should presumably be nonperturbative, so may provide nontrivial friction 
during the bubble expansion. 
In this sense the present analysis might merely give an ideal prediction to 
the created GW signals having its maximal size. 
More precise estimates might involve a quasi particle description for the 
dark gluonic plasma.

\item The universe may experience a stage of PYM dominance. In this scenario, the dark glueball has a relatively short lifetime, decaying away before the onset of BBN due to  the proper higher dimensional operators linking the two sectors. 

\item 
As can be read off from Table~\ref{table-2}, 
the parameter $\wt \beta$ decreases with the number of color $N$, implying that 
the PT in our models tends to happen more slowly, as $N$ gets larger. 
This tendency is still operative even starting from $N=3$. 
This decreasing $\tilde{\beta}$ in $N$ is a characteristic feature of our PLM, 
which is compared to the PT nature predicted from the matrix model~\cite{Halverson:2020xpg} exhibiting a growing scaling of the $\wt \beta$ in $N$   
for $N > 4$, but a decreasing scaling for $N<4$.    
At $N=4$ the parameters $\alpha$ and $\wt \beta$ have the same order of magnitude for both the matrix model and ours, so both models coincide at this $N=4$ 
and GW signals do as well. 
As one can see from Fig.~\ref{SM}, the peak strength of GW gets smaller as $N$ decreases,  
hence it will be impossible to detect for $N=3$ by the prospected detectors. 
Thus the GW detection for the larger $N$ signals would be a good 
discriminator between the matrix model and ours, where the latter's signals   
will be much stronger because of the much slower FOPTs. 
More detailed analysis along this line would be worth pursuing. 

\end{itemize}

\noindent 
\underline{\it Note added:} \\ 
\\
During the completion of this work, Ref.~\cite{Huang:2020mso} has been posted and discussed GW signals from dark $SU(N)$ PYM theory, also based on the PYM model with a Polynomial potential which includes all terms up to $|l|^8$. But we adopt a different strategy, considering the Polynomial potential merely including the 4-6 or 4-8 terms, motivated by the possible origin from integrating out a heavy PL with a higher center charge. We provide comparison between the two types of models. This point including the possible derivation of the 4-6 PLM has not been addressed in the literature.   Besides, our work shows that the PLMs with the Haar measure term are incompatible  with  the  large $N$ scaling  for the thermodynamic  quantities  and  the  latent  heat. 
\\

\noindent {\bf{Acknowledgements}}

We thank Bo Feng and Ruiwen Ouyang for useful discussions, and are grateful to Huaike Guo and Zhi-Wei Wang for useful comments.  
This work is supported in part by the National Science Foundation of China (11775086) and the National Key Research and Development Program of China Grant No.2020YFC2201504 (ZK). S.M. work was supported in part by the National Science Foundation of China (NSFC) under Grant No.11747308, 11975108, 12047569 and the Seeds Funding of Jilin University.

\end{document}